\documentclass[a4paper,fleqn,usenatbib]{mnras}

\usepackage{newtxtext,newtxmath}
\usepackage{enumerate}
\usepackage[T1]{fontenc}
\usepackage{ae,aecompl}
\usepackage{cleveref}

\usepackage{graphicx}	
\usepackage{amsmath}	
\usepackage{amssymb}	
\usepackage{color}

\title[Near infrared stellar populations in ETGs]{Probing Evolutionary Population Synthesis Models in the Near Infrared with Early Type Galaxies}

\author[Dahmer-Hahn et al.]{Luis Gabriel Dahmer-Hahn$^{1,2}$\thanks{E-mail: dahmer.hahn@ufrgs.br}, Rog\'erio Riffel$^{1}$, Alberto Rodr\'iguez-Ardila$^{3}$, \newauthor Lucimara P. Martins$^{4}$, Carolina Kehrig$^{5}$, Timothy M. Heckman$^{6}$, \newauthor Miriani G. Pastoriza$^{1}$, Natacha Z. Dametto$^{1}$.\\
$^{1}$Departamento de Astronomia, Universidade Federal do Rio Grande do Sul. Av. Bento Goncalves 9500, Porto Alegre, RS, Brazil.\\
$^{2}$Instituto de F\'isica e Qu\'imica, Universidade Federal de Itajub\'a, Brasil.\\
$^{3}$Laborat\'orio Nacional de Astrof\'isica - Rua dos Estados Unidos 154, Bairro das Na\c{c}\~oes. CEP 37504-364, Itajub\'a, MG, Brazil.\\
$^{4}$NAT - Universidade Cruzeiro do Sul, Rua Galv\~ao Bueno, 868, S\~ao Paulo, SP, Brazil.\\
$^{5}$Instituto de Astrof\'isica de Andaluc\'ia, CSIC, Apartado de correos 3004, 18080 Granada, Spain.\\
$^{6}$Center for Astrophysical Sciences, Department of Physics and Astronomy, The Johns Hopkins University, Baltimore, MD 21218, USA.\\
}
\date{Accepted XXX. Received YYY; in original form ZZZ}

\pubyear{2015}

\begin{document}
\label{firstpage}
\pagerange{\pageref{firstpage}--\pageref{lastpage}}
\maketitle

\begin{abstract}

  We performed a near-infrared (NIR, $\sim$1.0$\mu$m-2.4$\mu$m) stellar population study in a sample of early type galaxies. The synthesis was performed using five different evolutionary population synthesis libraries of models. Our main results can be summarized as follows: low spectral resolution libraries are not able to produce reliable results when applied to the NIR alone, with each library finding a different dominant population. The two newest higher resolution models, on the other hand, perform considerably better, finding consistent results to each other and to literature values. We also found that optical results are consistent with each other even for lower resolution models. We also compared optical and NIR results, and found out that lower resolution models tend to disagree in the optical and in the NIR, with higher fraction of young populations in the NIR and dust extinction $\sim$1 magnitude higher than optical values. For higher resolution models, optical and NIR results tend do aggree much better, suggesting that a higher spectral resolution is fundamental to improve the quality of the results.

\end{abstract}



\section{Introduction}

Understanding the processes involved in galaxy evolution is one of the main topics of modern astrophysics. These processes are mainly driven by the star formation history (SFH) of the galaxies. One of the main methods to access the unresolved stellar content of galaxies is by comparing their observed spectra with combinations of simple stellar population (SSPs) libraries. These libraries can be empirical (in this case being limited by the properties of nearby stellar clusters) or can be constructed by using knowledge about stellar evolution, a technique called evolutionary population synthesis \citep[EPS, e.g.][herefater BC03, M05, C09, MG15 and MIUSCAT, respectively]{BC03, M05, C09, MG15, rock16}. 

\par

The SSPs are usually constructed using one of the two alternatives: isochrone synthesis or ``fuel consumption based'' algorithms. With the first approach, SSPs are calculated by integrating the stellar contributions to the flux in the various pass-bands of all mass-bins along one isochrone, after assuming an Initial Mass Function (IMF, e.g. BC03). In the second approach, after leaving the main sequence, the duration of each subsequent phase in stellar evolution is calculated by using the fuel consumption theory (e.g. M05).

\par

The big problem is that both approaches result in very different luminosities for short evolutionary stages, especially the crucial TP-AGB phase, whose underlying physics is still poorly known \citep[M05,][]{marigo08, conroyEgunn10, conroy13, kriek+10, zibetti13, noel+13, riffel15}. This happens because some processes of stellar evolution (mass-loss, changing opacities, dredge-up events, etc.) are not well understood, and receive a different treatment in each model flavour (e.g. BC03, M05, C09, MG15 and MIUSCAT). Models based on the fuel consumption theory tend to overpredict TP-AGB features while those based on isochrone synthesis generally underestimate them \citep[][and references therein]{zibetti13}, although a few models based on isochrone synthesis also contain large amounts of these stars \citep[e.g.][]{marigo08}. Currently, there is no consensus regarding what TP-AGB contribution best reproduces the observed spectra of galaxies. Many stellar absorption features predicted by TP-AGB heavy models have been found, like the 1.1 $\mu m$ CN band \citep{riffel07}, the 1.4 $\mu m$ CN band \citep{martins13b} and the ZrO features at 0.8-1.0 $\mu m$ \citep{martins13b}. On the other hand, \citet{zibetti13} did not detect the TP-AGB spectral features predicted by M05 in their spectra of post-starburst galaxies at z$\sim$0.2. \citet{riffel15} found that models based on empirical libraries that predict relatively strong near infrared (NIR) features provide a more accurate description of the data. However, none of the models tested by them successfully reproduces all of the features observed in the spectra. Also, \citet{riffel15} claimed that stars in other evolutionary phases like RGB may be crucial to describe the absorption features detected in galaxies and that empirical spectra of these kind of stars should be included in the EPS models. 

\par

The libraries of SSPs are then used by a computing code \citep[e.g. {\sc starlight,}][]{CF05} to determine parameters such as ages, element abundances, stellar masses and stellar mass functions by searching for the combination of SSPs that best reproduces the observed spectrum.
\par
A major issue when characterizing the stellar population of a galaxy is atributed to the dependence of results in the model set used when fitting the underlying stellar features \citep{chen10}. This happens because of all the uncertainties related to the construction of the SSPs. Summed to these effects, there is the well known age-metallicity degeneracy, which difficults to distinguish between an old stellar population and a reddened or more metallic younger one \citep{worthey94}.

\par

Here we aim to compare, for the first time in the literature, the stellar population predictions derived using NIR spectra of different sets of EPS models for the central region of 6 local early-type galaxies (ETGs) and one spiral galaxy. The former morphologycal type was chosen because it contains a relatively homogeneous old stellar population. We are aware, though, that current day interpretation is that these objects experienced moderately different star formation histories, with the present-day stellar populations slightly differing in metallicity and/or age \citep[][and references therein]{rickes09}.

\par

This paper is structured as follows: The data and reduction process are presented in Section~\ref{sec:data}. The Stellar Population Synthesis method used in the analysis is presented in Section~\ref{sec:synth}. In Section~\ref{sec:results}, we present the main results from the synthesis using the different sets of SSPs. A discussion of these results is in Section~\ref{sec:discussion}. The final remarks are given in section ~\ref{sec:finalremarks}.

\section{Data and reduction}
\label{sec:data}

From the 6 ETGs, 5 are from the Calar Alto Legacy Integral Field Area Survey \citep[CALIFA][]{sanchez+12, husemann+14, sanchez+16}. These objects were selected for being the ETGs in the CALIFA data release 1 and accesible to the night sky during the observing run. Also, the CALIFA papers confirmed the presence of warm gas for these 5 galaxies. The final 7 targets were selected on the basis of surface brightness to offer a good compromise between S/N and exposure time. We also observed NGC 4636, a typical LINER, and NGC 5905, classified as SB(r)b, for comparison. The infrared data were obtained at the Astrophysical Research Consortium (ARC) telescope. The TripleSpec \citep{wilson04} instrument was used to obtain cross-dispersed spectra in the range $0.95 - 2.46 \mu m$. We used the 1.1 $''$ slit, resulting in a spectral resolution of R$\sim$2000. After each target, we observed an A0V, A1V or A2V star at a similar airmass for flux calibration and telluric correction. Both the science objects and the telluric stars were observed following the dithering pattern object-sky-object. Internal flat-field and arc lamp exposures were  also acquired for  pixel response  and wavelength  calibration,  respectively. The reduction of the data was done using Triplespectool, a modified version of Spextool \citep{vacca04,cushing04} using standard settings. These spectra are available for download at the MNRAS website.

\par

\begin{figure*}
        \noindent
	\includegraphics[width=\textwidth]{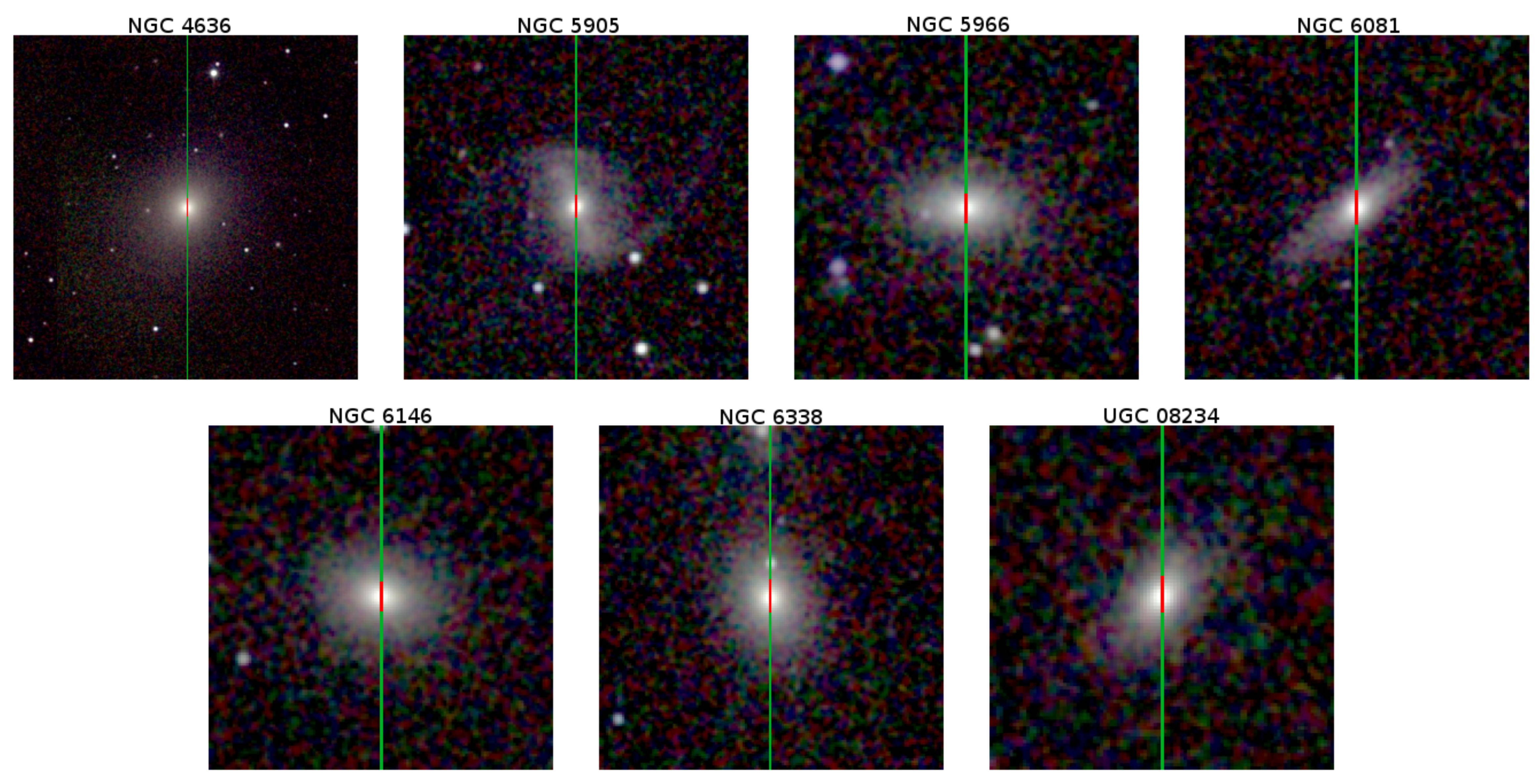}
    \caption{Combined 2MASS JHKs imaging  for the galaxy sample. The vertical green line shows the slit orientation while the red segment indicates the aperture used to extract the spectra.}
    \label{fig:figsample}
\end{figure*}

\begin{figure}
        \noindent
	\includegraphics[width=\linewidth]{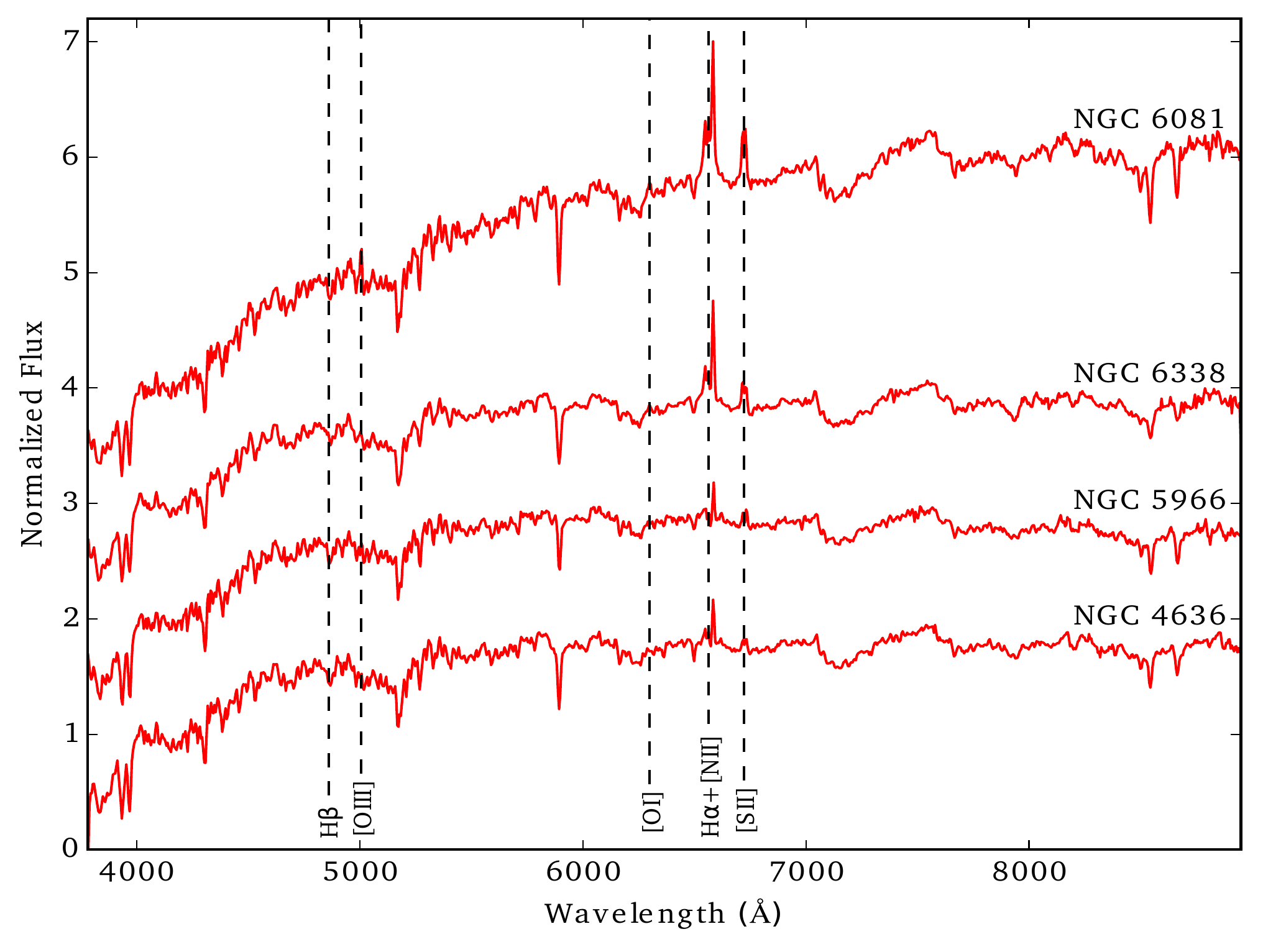}
    \caption{Red curves represent SDSS spectra, which are available for four galaxies of the sample. Rest-frame wavelengths of the main optical emission-lines are indicated.}
    \label{fig:opticalspec}
\end{figure}

\begin{figure*}
        \noindent
	\includegraphics[width=\textwidth]{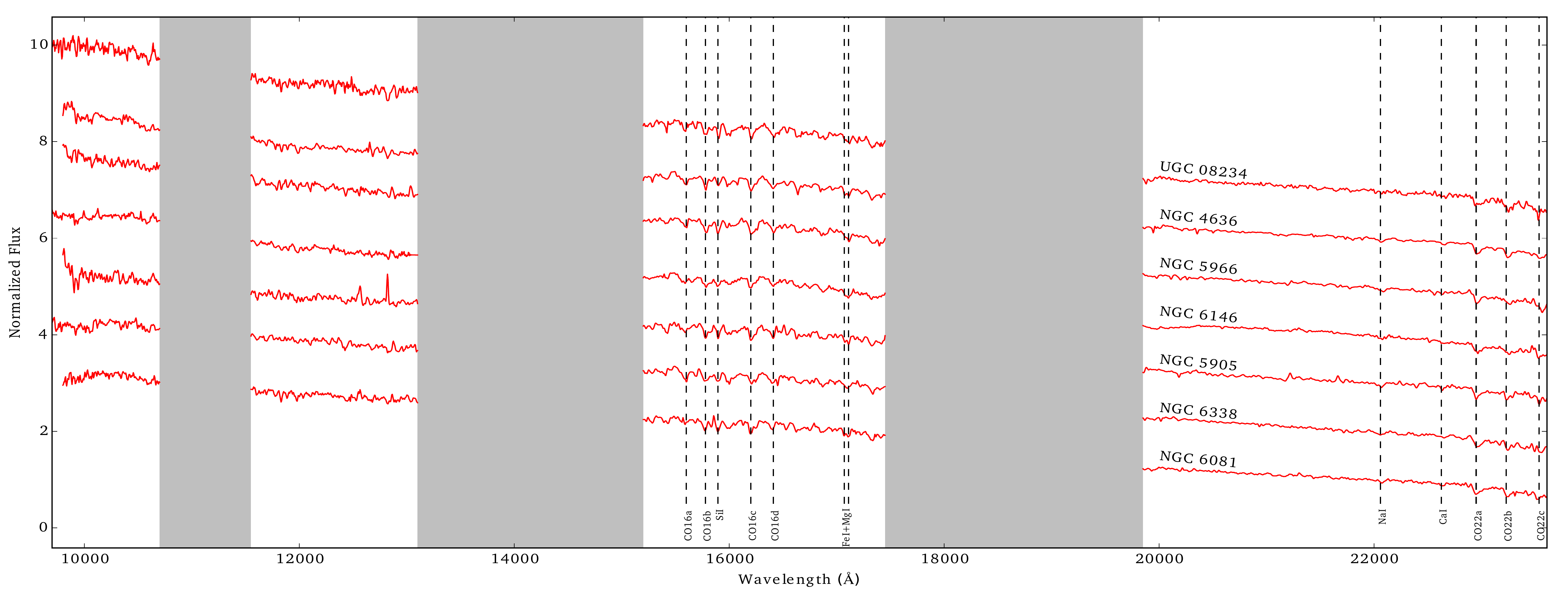}
    \caption{NIR spectra for the sample. The main absorption bands are indicated by dashed lines. High telluric absorption regions are shaded.}
    \label{fig:nirspec}
\end{figure*}

Table~\ref{tab:tabsample} shows the basic properties for the sample. Also, in order to compare NIR results with optical ones, for the five CALIFA objects, we extracted optical spectra using apertures of size similar to that of the NIR slit. For NGC 4636, which is not within the CALIFA targets, we used optical spectra from SDSS. Figure~\ref{fig:figsample} shows Two Micron All Sky survey (2MASS) JHK imaging of the sample with the slit orientation. The individual description of the objects of the sample is presented in the Appendix~\ref{sec:sample}. Optical and NIR spectra are shown in Figure~\ref{fig:opticalspec} and Figure~\ref{fig:nirspec}, respectively.

\begin{table*}
	\centering
	\caption{Basic properties for the sample.}
	\label{tab:tabsample}
	\begin{tabular}{lcccccccr}
		\hline
		Object   & Morphology $^1$ & K magnitude $^2$ & Absolute   & z $^4$   & Aperture Radius & Exposure Time & S/N$^5$ &Optical$^6$  \\
		         &                 &                 & K magnitude $^3$ &     &    (arcsec)     &        (min)    &         & spectrum   \\
		\hline
		NGC 4636 & E0-1            & 9.0             &  -21.6      & 0.003129 & 10.0           & 27.0           & 54      & SDSS\\
		NGC 5905 & SB(r)b          & 11.0            &  -22.4      & 0.011308 & 7.0            & 24.0           & 30      & ---\\
		NGC 5966 & E               & 10.7            &  -23.3      & 0.014924 & 6.0            & 24.0           & 24      & CALIFA\\
		NGC 6081 & S0              & 10.5            &  -23.8      & 0.017265 & 6.0            & 36.0           & 45      & CALIFA\\
		NGC 6146 & E?              & 10.3            &  -25.2      & 0.029420 & 3.0            & 36.0           & 56      & CALIFA\\
		NGC 6338 & S0              & 10.6            &  -24.7      & 0.027427 & 4.0            & 36.0           & 30      & CALIFA\\
		UGC 08234 & S0/a           & 10.7            &  -24.6      & 0.027025 & 6.0            & 36.0           & 32      & CALIFA\\
		\hline
	\end{tabular}
	\begin{list}{Table Notes:}
	\item $^1$ \citet{devaucouleurs91} $^2$ Two Micron All Sky survey team, 2003, 2MASS extended objects, final release $^3$ Calculated based in the apparent magnitude and the redshift and assuming H=70 km\,$s^{-1}$\,Mpc$^{-1}$ $^4$ NED $^5$ K-band Signal-to-Noise ratio before smoothing the spectra. $^6$ Presence or absence of SDSS spectra.
	\end{list}
\end{table*}

\section{Stellar Population Synthesis}
\label{sec:synth}

The stellar population synthesis technique consists basically of comparing the observed spectrum of a galaxy with a combination of SSPs with different ages and metallicities \citep[this set of SSPs is known as a base of elements,][]{CF05}, searching for a combination of SSPs that suitably fits the observed spectrum.

\par

For a proper fitting of a galaxy's stellar population, the library of models must cover the range of possible observed spectral properties (e.g. ages and metallicities). It is also fundamental to choose an adequate number of elements for such library in order to have non-degenerate solutions \citep{schmidt91,CF05,dametto14}. \citet{chen10}, presented a study in the optical region showing that different models may result in quite different SFHs for the same observed spectrum. Studying six different types of galaxies (star-forming galaxies, composite galaxies, Seyfert 2s, LINERs, E+A, and early-type galaxies) using 6 different EPS models, they found that the differences found are significant, but the dominant populations are unaltered. Also, they found that using the same models, the results depend on the selected ages. In the NIR, this scenario seems to be more dramatic, since the inclusion of the TP-AGB phase in the models is still very uncertain and is a matter of debate \citep[M05,][]{marigo08, conroyEgunn10, conroy13, kriek+10, zibetti13, noel+13, riffel15}. To properly address this issue we decided to build eight different libraries of models using five different EPS models flavours, as follows:

\begin{enumerate}[i.]
  \item \citet[BC03]{BC03}
  \item \citet[M05]{M05}
  \item \citet[C09]{C09}
  \item \citet[MG15]{MG15}
  \item \citet[MIUSCAT]{rock16}
\end{enumerate}

Details of the EPS models, as well as the choosen ages, metallicities and evolutionary tracks of the  SSPs used in the fitting are listed in Table~\ref{tab:SSPs}. Since MG15 and MIUSCAT have only SSPs with ages t$\geq$1Gyr, in order to allow for a suitable comparison of the results from the different EPS models, we created 3 additional libraries of models by removing the SSPs with ages t<1 Gyr from BC03, M05 and C09 models. These libraries do not include SSPs younger than 1Gyr because of the lack of hot stars in the IRTF library \citep{cushing05,rayner09}, which was used to build the models.

\par

It is worth mentioning that in the NIR, the spectral resolution of BC03, M05 and C09 is lower than that of the observed spectra. We then rebinned the data in order to match the spectral resolution of the models. Considering that MG15 and MIUSCAT have a spectral resolution similar to that of the observed spectra, this procedure was not necessary when these later libraries were employed.

\par

Regarding TP-AGB teatment, BC03 uses low-resolution stellar templates from \citet{hofner+00}, while M05 and C09 use higher resolution TP-AGB spectra from \citet{lanconEmouhcine02}, degrading them to match the rest of the spectra. MG15 and MIUSCAT use the IRTF library, which also contains TP-AGB stars \citep{riffel15}.

\begin{table*}
        \centering
        \tiny
	\caption{Basic properties for the EPS models and informations on the used SSPs.}
	\label{tab:SSPs}
	\begin{tabular}{lcccccc}
		\hline
		SSP & Ages (Gyr) & Metallicities ($Z\sun$) & Spectral resolution &Stellar & IMF & Evolutionary\\
                & & & (Optical/NIR) & Library & & Track\\
		\hline
		& 0.001, 0.0031, 0.0050, 0.0066, 0.0087, \\ & 0.01, 0.014, 0.025, 0.04, 0.055, \\BC03 & 0.1, 0.16, 0.28, 0.50, 0.9, & 0.005, 0.02, 0.2, 0.4, 1, 2.5 & 2000/300 & STELIB$^1$/& Salpeter$^3$ & Padova$^4$\\ & 1.27, 1.43, 2.5, 4.25, 6.25, & & & BaSeL 3.1$^2$\\ & 7.5, 10, 13 and 15\\ \hline
                BC03io & 1.27, 1.43, 2.5, 4.25, 6.25,& 0.005, 0.02, 0.2, 0.4, 1, 2.5 & 2000/300 & STELIB/& Salpeter & Padova\\ & 7.5, 10, 13 and 15 & & & BaSeL 3.1\\ \hline
                & 0.001, 0.003, 0.0035, 0.004, 0.005, 0.0055,\\ & 0.006, 0.0065, 0.007, 0.0075, 0.008,\\ M05 & 0.0085, 0.009, 0.01, 0.015, 0.02, & 0.02, 0.5, 1 and 2 & 300/300 & BaSeL 2.2$^2$ & Salpeter & Cassisi$^5$ $+$\\ & 0.025, 0.03, 0.05, 0.08, 0.2, & & & & & Geneva$^6$\\ & 0.03, 0.04, 0.05, 0.07, 0.08,\\ & 1.0, 1.5, 2.0, 3.0 and 13 \\ \hline
                M05io & 1.0, 1.5, 2.0, 3.0 and 13 & 0.02, 0.5, 1 and 2 & 300/300 & BaSeL 2.2 & Salpeter & Cassisi $+$\\ & & & & & & Geneva\\\hline
                & 0.00032, 0.00075, 0.0025, 0.0035, 0.004,\\ & 0.0044, 0.005, 0.0056, 0.0063, 0.0075,\\ C09 & 0.0079, 0.0084, 0.0089, 0.016, 0.032, & 0.02, 0.3, 0.95 and 1.5 & 300/300 & BaSeL 3.1 & Salpeter & Padova\\ & 0.038, 0.045, 0.063, 0.071, 0.089,\\ & 0.21, 0.32, 0.42, 0.53, 0.89,\\ & 2.11, 5.31, 6.31, 8.91 and 12.59\\\hline
		C09io & 2.11, 5.31, 6.31, 8.91 and 12.59 & 0.02, 0.3, 0.95 and 1.5 & 300/300 & BaSeL 3.1 & Salpeter & Padova\\\hline
                & 1.0, 1.12, 1.26, 1.41, 1.58, 1.78, 2.0, 2.24, 2.51, \\ MG15 & 2.82, 3.16, 3.55, 3.98, 4.47, 5, 5.62, 6.31, & 0.0038, 0.0076, 0.019 and 0.03 & ---/2000 & IRTF$^{7}$ & Salpeter & Padova\\ &7, 7.94, 8.91, 10, 11.2, 12.6 and 14.1\\\hline
                & 1.0, 1.25, 1.5, 1.75, 2.0, 2.25, 2.5, 2.75,
                \\ MIUSCAT & 3.0, 3.25, 3.5, 3.75, 4.0, 4.5, 5.0, 5.5, 6.0, & 0.0085, 0.01, 0.022, 0.027 & ---/2000 & IRTF & Revised & BaSTI$^{9}$\\
                & 6.5, 7.0, 7.5, 8.0, 8.5, 9.0, 9.5, 10.0, 10.5, &&&&Kroupa$^{8}$\\ & 11.0, 11.5, 12.0, 12.5, 13.0, 13.5 and 14.0 \\\hline
		\hline
	\end{tabular}
        \begin{list}{Table Notes:}
        \item (1) \citet{lebogne03} (2) \citet{lejeune97,lejeune98,westera02} (3) \citet{salpeter} (4) \citet{marigo08} and references therein (5) \citet{cassisi97a,cassisi97b} (6) \citet{schaller92} (7) \citet{cushing05,rayner09} (8) \citet{kroupa01} (9) \citet{BaSTI}
        \end{list}
\end{table*}

\par

Since our objects may host a low-luminosity AGN (LLAGN), we followed \citet[][and references therein]{riffel09}. In addition, we added to the base set a power law with $F_\nu \propto \nu^{-1.5}$ to represent the featureless continuum (FC) emission from the AGN. Moreover, in the NIR, 8 planck distributions with temperatures from 700 to 1400 K were also added, in steps of 100 K. They represent hot dust continuum heated by the AGN \citep[see][]{riffel09}. 

\subsection{Fitting Procedures}
\label{sec:method}

In order to fit the stellar populations, we used the {\sc starlight} code, which is described in \citet{CF04, CF05}. In summary, the code fits the observed spectrum with a combination of SSPs in different proportions. It also searches for the internal extinction that best describes the observed spectrum. To this aim, we used the \citet{ccm} reddening law. Basically, by using a $\chi^2$ minimization approach, {\sc starlight} fits an observed spectrum $O_{\lambda}$ with a combination of $N_{\star}$ SSPs solving the equation:

$$M_\lambda = M_{\lambda 0} [\sum_{j=1}^{N\star} x_j b_{j,\lambda} r_\lambda] \otimes G(v_\star, \sigma_\star)$$

\noindent
where $M_{\lambda}$ is a model spectrum, $M_{\lambda_0}$ is the flux at the normalized point $\lambda_0$, $N_\star$ is the number of SSPs used to compose the model, $\vec x$ is the population vector so that $x_j$ indicates the contribution from the j-esim SSP normalized at $\lambda_0$, $b_{j,\lambda}$ is the j-esim model spectrum, $r_\lambda$ is the reddening factor $r_\lambda=10^{-0.4(A\lambda-A\lambda_0)}$, which is parameterized by the dust extinction, Av. The stellar velocity dispersions and group velocities are modelled by a gaussian function $G(v_\star, \sigma_\star)$.

\section{Results}
\label{sec:results}

\begin{figure}
        \noindent
	\includegraphics[width=\linewidth]{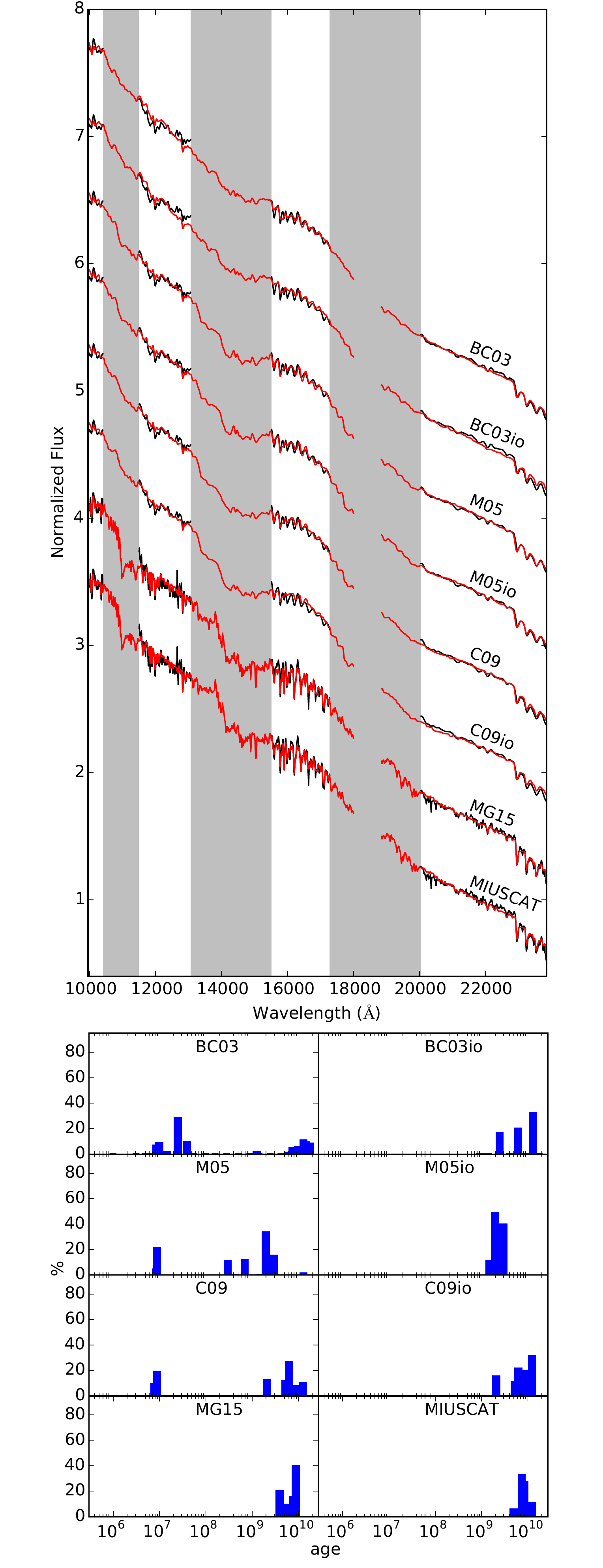}
    \caption{NIR results for NGC 4636 using the 8 libraries of models. On the upper pannel, observed spectrum is shown in black and modelled spectrum is shown in red. Masked areas are shaded. On the bottom pannel, we show in blue the percentage luminosity contribution for every age.}
    \label{fig:4636}
\end{figure}

\begin{figure}
        \noindent
	\includegraphics[width=\linewidth]{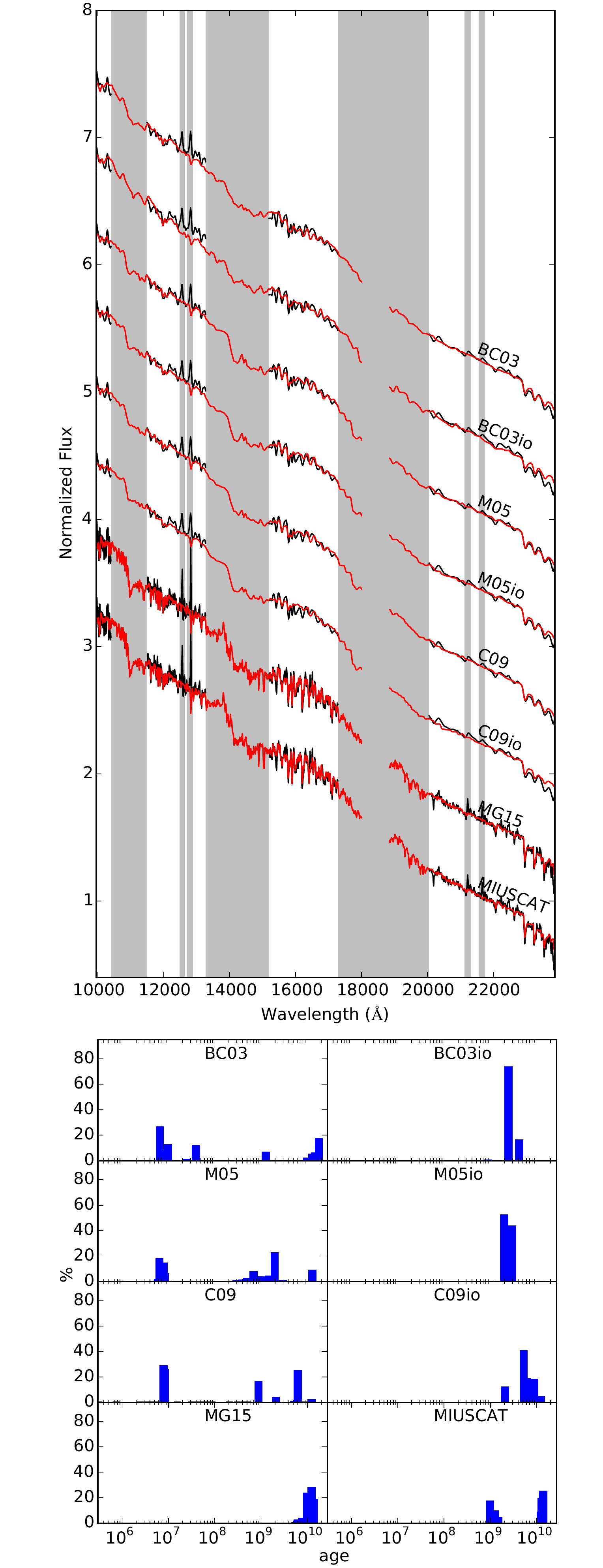}
    \caption{Same as Figure~\ref{fig:4636} for NGC 5905. Emission lines are also shaded on the upper panel.}
    \label{fig:5905}
\end{figure}

\begin{figure}
        \noindent
	\includegraphics[width=\linewidth]{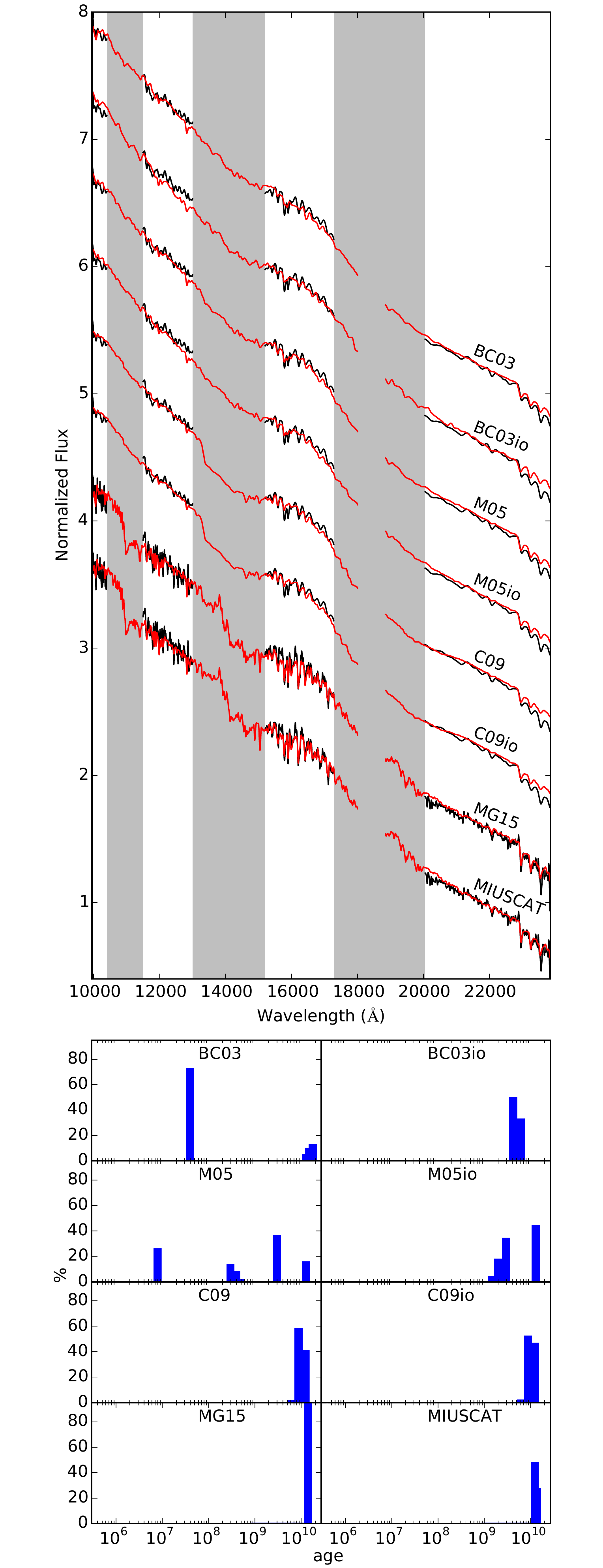}
    \caption{Same as Figure~\ref{fig:4636} but for NGC 5966}
    \label{fig:5966}
\end{figure}

\begin{figure}
        \noindent
	\includegraphics[width=\linewidth]{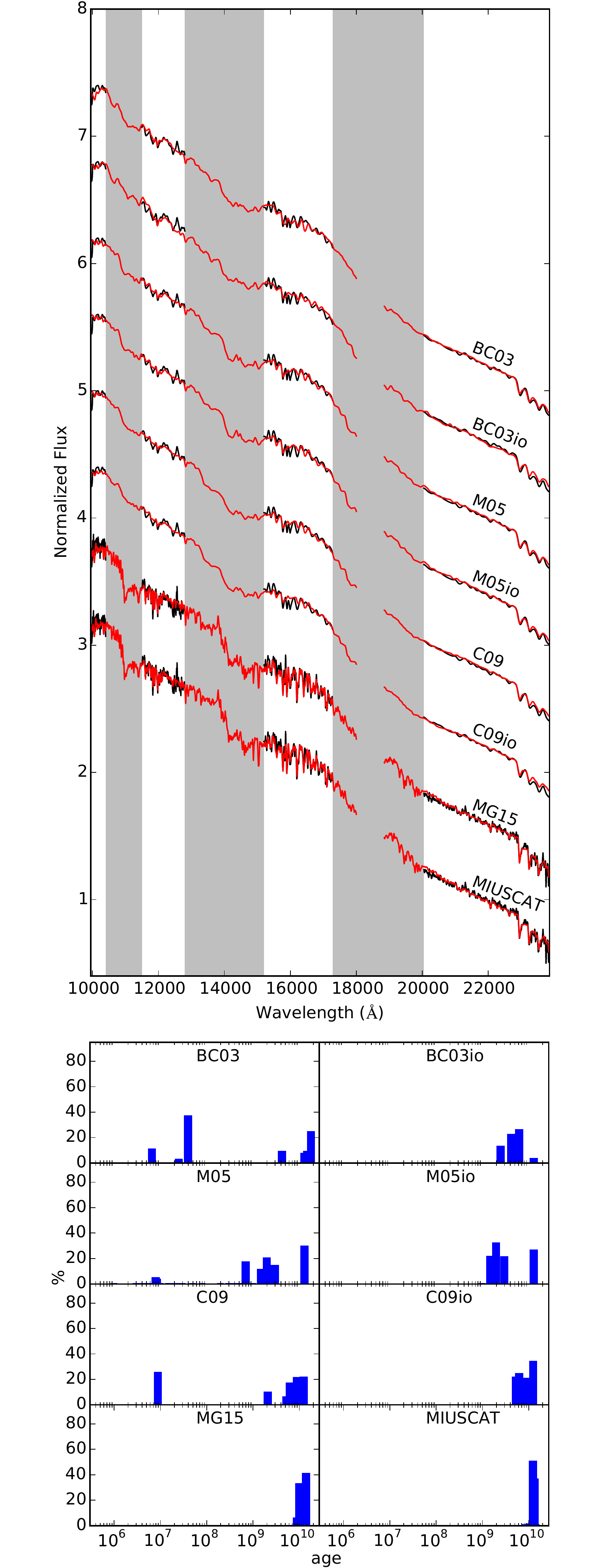}
    \caption{Same as Figure~\ref{fig:4636} but for NGC 6081}
    \label{fig:6081}
\end{figure}

\begin{figure}
        \noindent
	\includegraphics[width=\linewidth]{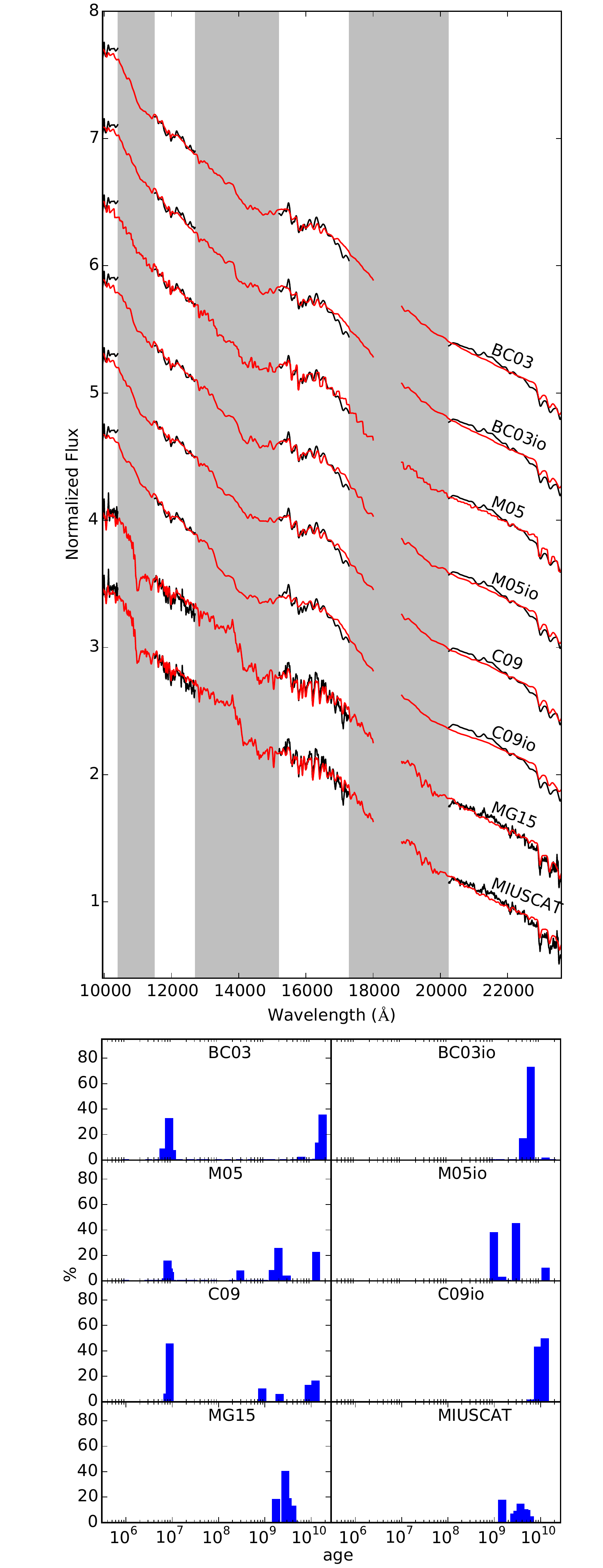}
    \caption{Same as Figure~\ref{fig:4636} but for NGC 6146}
    \label{fig:6146}
\end{figure}

\begin{figure}
        \noindent
	\includegraphics[width=\linewidth]{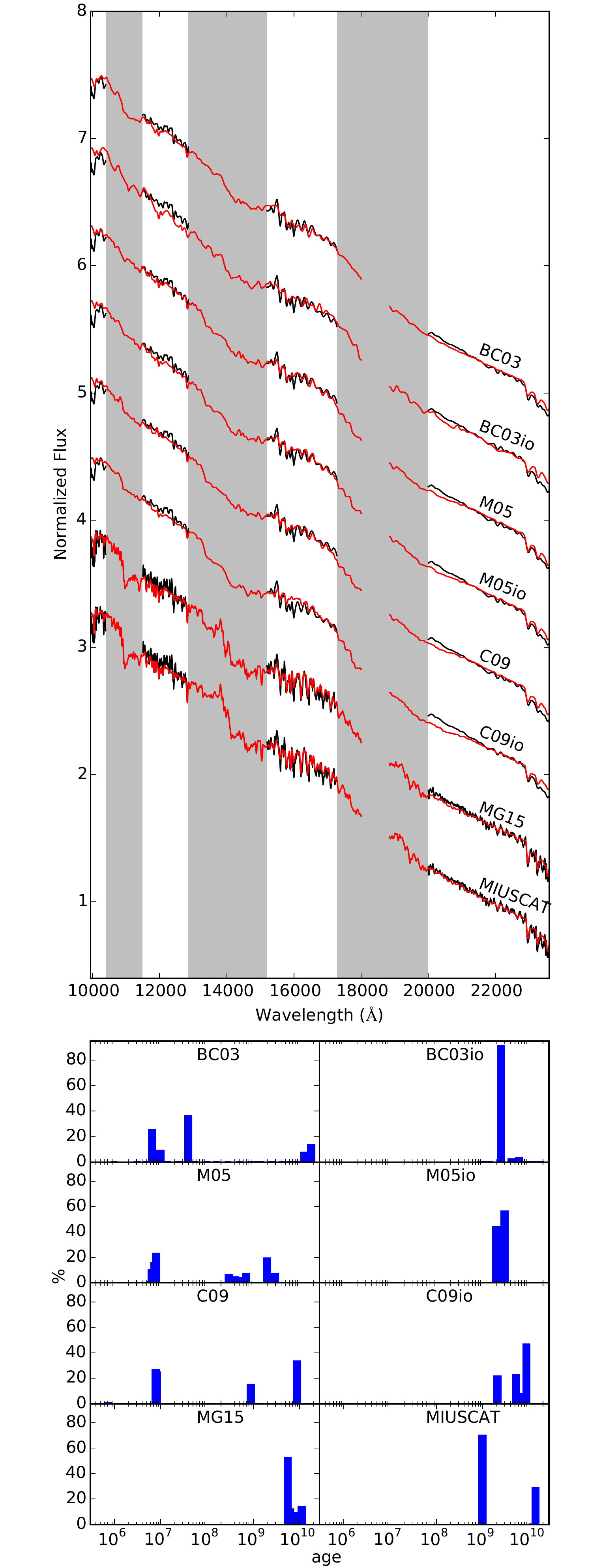}
    \caption{Same as Figure~\ref{fig:4636} but for NGC 6338}
    \label{fig:6338}
\end{figure}

\begin{figure}
        \noindent
	\includegraphics[width=\linewidth]{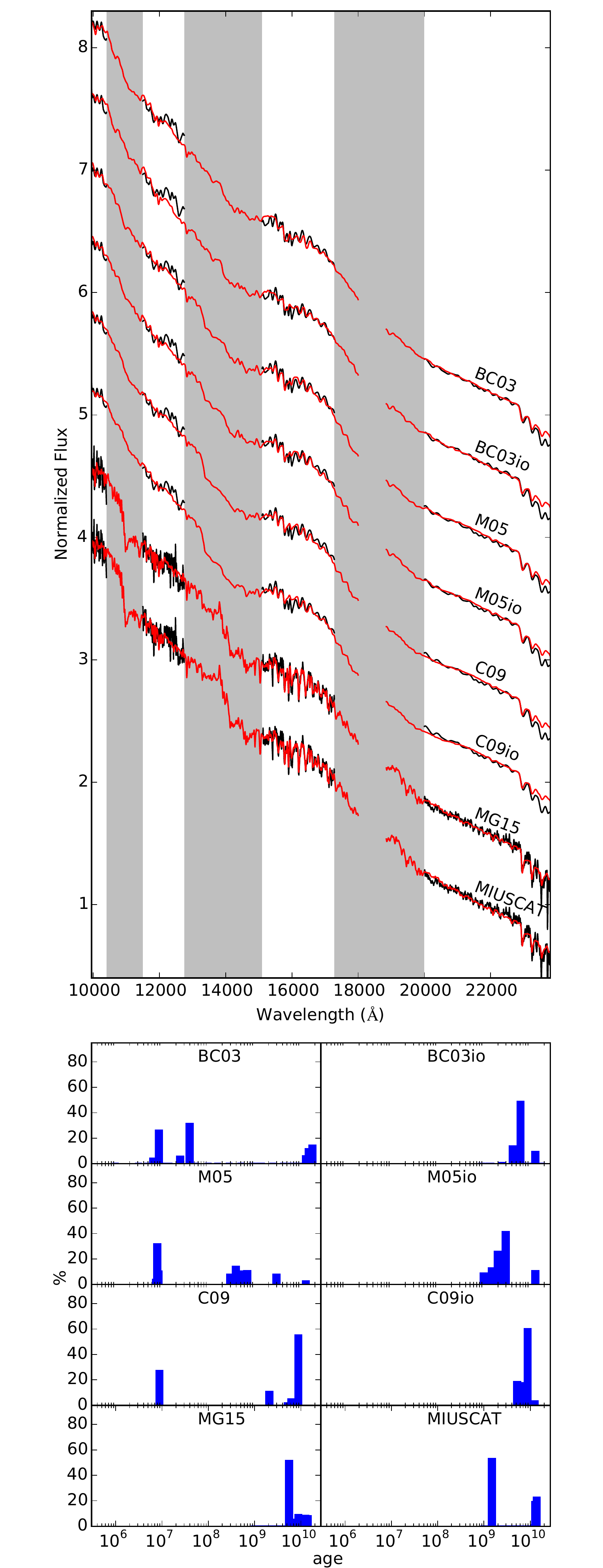}
    \caption{Same as Figure~\ref{fig:4636} but for UGC 08234}
    \label{fig:08234}
\end{figure}

The upper pannels of \cref{fig:4636,fig:5905,fig:5966,fig:6081,fig:6146,fig:6338,fig:08234} show, for each galaxy of the sample, the observed (black) and modeled (red) spectra for each of the 8 libraries of models employed. Areas with high atmospheric absortion were shaded in the Figures. The bottom pannels show in blue the luminosity contributions (metallicities summed) for every age of the best model fitted by {\sc starlight}.

\par

Since small differences in the ages of the stellar populations are washed away by the noise present in real data, a more consistent and robust way to present the results is in the form of condensed population vectors \citep{CF04, CF05}. With this in mind, we followed \citet{riffel09} and defined the light fraction population vectors as follows: {\it xy} (t $\leqslant$ 50Myr), {\it xi} (50Myr $<$ t $\leqslant$ 2Gyr) and {\it xo}(t $>$ 2Gyr) to represent the young, intermediate and old stellar population vectors, respectively. The same age bins were used for the mass fraction population vectors. Besides, we defined cold and hot dust emission vectors, $BB_c$ for T $\leqslant$ 1000K and $BB_h$ for T $>$ 1000K. The results for these binned population vectors for each galaxy are presented in Table~\ref{tab:NIRtable}.

\par

\begin{figure}
        \noindent
	\includegraphics[width=\linewidth]{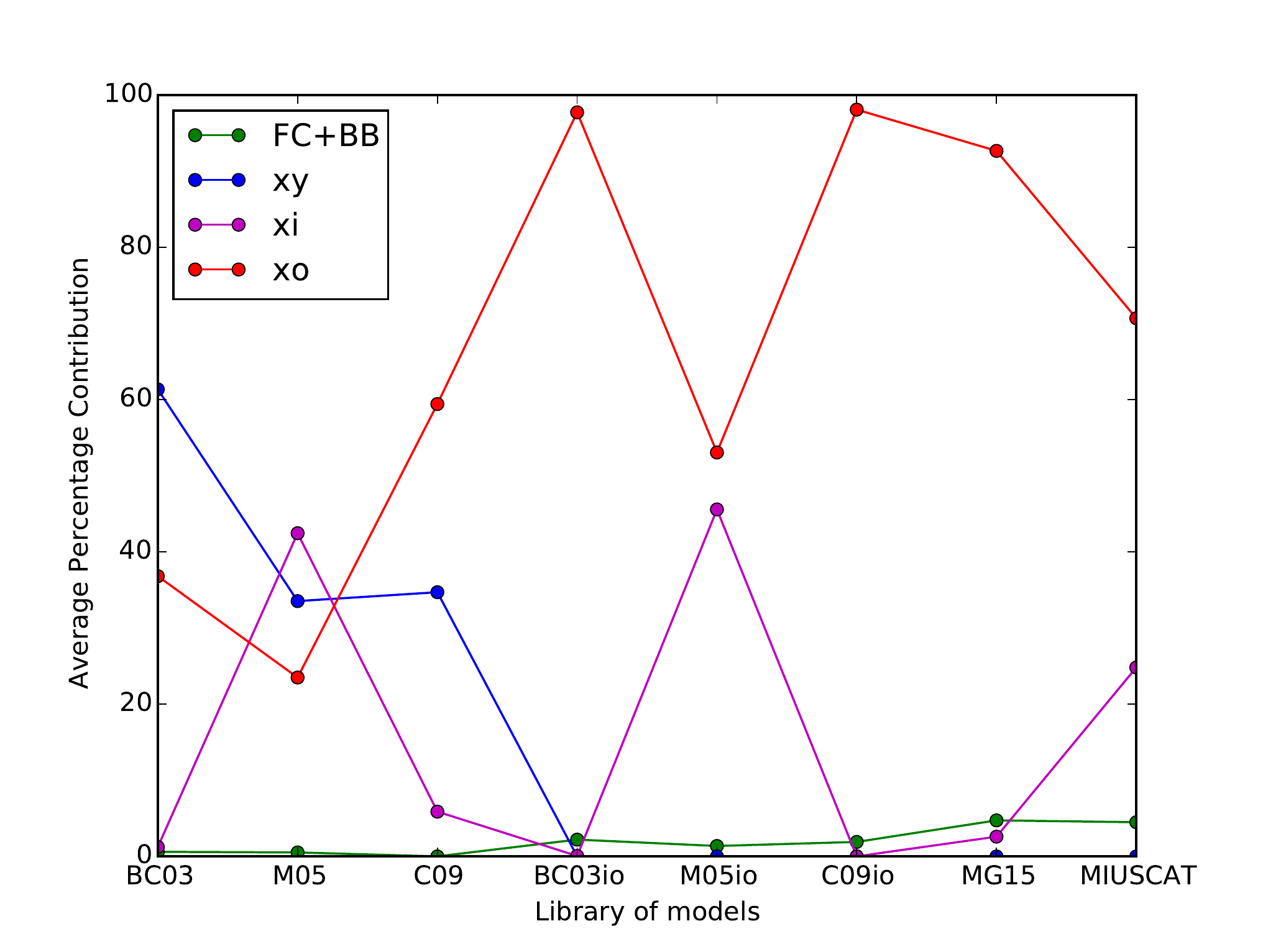}
    \caption{Average percentage contribution for each library of models. Young populations are displayed in blue, intermediate-aged populations in magenta and old stellar populations in red. In green, the summed contributions from featureless continuums and black bodies.}
    \label{fig:percentage}
\end{figure}

We also followed \citet{CF05}, who proposed two additional parameters to describe the SP mixture of a galaxy. These parameters are the mean stellar age (\textit{<t>}) and mean metallicity (\textit{<Z>}), which are defined by the following equations
$$<t>_L = \sum_{j=1}^{N\star}x_j log(t_j) $$
$$<Z>_L = \sum_{j=1}^{N\star}x_j Z_j $$
where $t_j$ and $Z_j$ are the age and metallicity of the j-esim SSP. The $x_j$ percentage contribution can be weighted by light (L) and mass (M) fractions. These parameters, together with the Adev, are also listed in Table~\ref{tab:NIRtable} (Adev is a value that measures the fit quality, according to the relation $|O_\lambda - M_\lambda|/O_\lambda$). For a better visualization of the trends on each library of models, we summed the percentage contribution of the 7 objects and presented them on Fig~\ref{fig:percentage}.

\begin{table*}
\scriptsize
\centering
\caption{Results for the NIR synthesis.}
\label{tab:NIRtable}
\begin{tabular}{lccccccccccccr}
\hline
Library	              &FC1.50   &BB$_c$     &BB$_h$   &xy         &xi        &xo        &my      &mi        &mo        &Av        &$<t>_L$    &$<Z>_L$  &Adev\\
\hline
&&&&&&&NGC 4636\\
\hline
BC03                  &0.0     &0.0      &0.0    &56.8      &2.0      &41.0     &5.7    &0.8      &93.4     &1.52      &8.49       &0.01796  &1.34\\
M05                   &0.0     &0.0      &0.0    &26.1      &57.1     &16.7     &1.4    &51.1     &47.3     &1.12      &8.58       &0.02362  &1.05\\
C09                   &0.0     &0.0      &0.0    &30.1      &0.0      &69.9     &---    &---      &---      &1.33      &8.91       &0.01439  &1.27\\
BC03io                &0.0     &0.0      &0.0    &0.0       &0.0      &100.0    &0.0    &0.0      &100.0    &1.23      &9.90       &0.01679  &1.86\\
M05io                 &0.0     &0.0      &0.0    &0.0       &60.2     &39.7     &0.0    &40.0     &59.9     &1.02      &9.36       &0.02579  &1.08\\
C09io                 &0.0     &0.0      &0.0    &0.0       &0.0      &100.0    &---    &---      &---      &1.11      &9.84       &0.009632 &1.54\\
MG15                  &0.0     &0.0      &6.4    &0.0       &0.0      &93.5     &---    &---      &---      &0.27      &9.82       &0.02089  &1.50\\
MIUSCAT               &0.0     &0.0	 &6.7    &0.0       &0.0      &93.2     &---    &---      &---      &0.57      &9.91       &0.01624  &1.85\\
\hline
&&&&&&&NGC 5905\\
\hline
BC03                  &4.2     &0.0      &0.0    &59.4      &6.6      &29.7     &4.4    &3.0      &92.5     &2.12      &8.16       &0.02282  &1.28\\
M05                   &3.6     &0.0      &0.0    &43.7      &43.7     &8.8      &3.3    &39.9     &56.7     &1.96      &8.18       &0.02438  &1.02\\
C09                   &0.0     &0.0      &0.0    &53.4      &16.1     &30.4     &---     &---     &---      &2.38      &8.11       &0.02211  &1.08\\
BC03io                &8.2     &0.0      &0.0    &0.0       &0.3      &91.3     &0.0    &0.4      &99.5     &1.43      &9.44       &0.02104  &1.85\\
M05io                 &5.6     &0.0      &0.0    &0.0       &51.5     &42.8     &0.0    &34.9     &65.0     &1.58      &9.38       &0.02081  &1.21\\
C09io                 &7.7     &0.0      &0.0    &0.0       &0.0      &92.2     &---     &---     &---      &1.74      &9.75       &0.006616 &1.56\\
MG15                  &13.3    &0.0      &0.0    &0.0       &0.0      &86.6     &---     &---     &---      &0.93      &10.06      &0.005964 &1.59\\
MIUSCAT               &4.5     &4.0	 &0.0    &0.0       &31.0     &60.3     &---     &---     &---      &1.10      &9.76       &0.01444  &1.63\\
\hline
&&&&&&&NGC 5966\\
\hline
BC03                  &0.0     &0.0      &0.0    &72.5      &0.0      &27.4     &12.5   &0.0      &87.5     &1.70      &8.31       &0.0146   &1.85\\
M05                   &0.0     &0.0      &0.0    &25.4      &23.3     &51.2     &0.8    &4.5      &94.6     &1.54      &8.70       &0.01968  &2.12\\
C09                   &0.0     &0.0      &0.0    &0.0       &0.0      &100.0    &---     &---     &---      &1.41      &10.01      &0.002981 &2.02\\
BC03io                &0.0     &0.0      &0.0    &0.0       &0.0      &100.0    &0.0    &0.0      &100.0    &1.29      &9.78       &0.01137  &2.54\\
M05io                 &0.0     &0.0      &0.0    &0.0       &21.6     &78.3     &0.0    &6.2      &93.7     &1.41      &9.71       &0.0157   &2.20\\
C09io                 &0.0     &0.0      &0.0    &0.0       &0.0      &100.0    &---     &---     &---      &1.42      &10.02      &0.003037 &2.01\\
MG15                  &0.0     &0.0      &0.0    &0.0       &0.0      &100.0    &---     &---     &---      &0.77      &10.15      &0.006124 &2.11\\
MIUSCAT               &0.0     &0.0	 &0.0    &0.0       &0.0      &100.0    &---     &---     &---      &0.53      &10.12      &0.02184  &2.01\\
\hline
&&&&&&&NGC 6081\\
\hline
BC03                  &0.0     &0.0      &0.0    &50.2      &0.0      &49.7     &4.3    &0.0      &95.6     &1.85      &8.76       &0.02432  &0.94\\
M05                   &0.0     &0.0      &0.0    &8.2       &48.1     &43.6     &0.2    &19.2     &80.5     &1.66      &9.28       &0.02499  &0.97\\
C09                   &0.0     &0.0      &0.0    &25.0      &0.0      &74.9     &---    &---      &---      &1.87      &9.13       &0.01255  &1.04\\
BC03io                &0.0     &0.0      &0.0    &0.0       &0.0      &100.0    &0.0    &0.0      &100.0    &1.62      &9.85       &0.02028  &1.28\\
M05io                 &0.0     &0.0      &0.0    &0.0       &52.9     &47.0     &0.0    &23.8     &76.2     &1.61      &9.52       &0.02437  &1.00\\
C09io                 &0.0     &0.0      &0.0    &0.0       &0.0      &100.0    &---     &---     &---      &1.80      &9.92       &0.01291  &1.06\\
MG15                  &0.0     &2.3      &0.0    &0.0       &0.0      &97.6     &---     &---     &---      &1.04      &10.07      &0.0131   &1.31\\
MIUSCAT               &0.0     &3.6	 &0.0    &0.0       &0.0      &96.3     &---     &---     &---      &0.90      &10.11      &0.02622  &1.34\\
\hline
&&&&&&&NGC 6146\\
\hline
BC03                  &0.0     &0.0      &0.0    &48.7      &0.0      &51.2     &1.1    &0.0      &98.8     &1.28      &8.61       &0.01962  &1.88\\
M05                   &0.0     &0.0      &0.0    &32.6      &41.1     &26.2     &0.7    &20.5     &78.7     &1.00      &8.64       &0.03     &1.73\\
C09                   &0.0     &0.0      &0.0    &55.3      &9.9      &34.7     &---    &---      &---      &1.42      &8.17       &0.02094  &1.72\\
BC03io                &7.0     &0.0      &0.1    &0.0       &0.0      &92.8     &0.0    &0.0      &100.0    &0.75      &9.77       &0.01976  &2.15\\
M05io                 &0.0     &0.0      &3.9    &0.0       &40.8     &55.2     &0.0    &13.9     &86.0     &0.77      &9.34       &0.03544  &1.77\\
C09io                 &0.0     &0.0      &5.5    &0.0       &0.0      &94.4     &---    &---      &---      &0.88      &10.03      &0.01448  &2.30\\
MG15                  &0.0     &0.0      &8.4    &0.0       &18.1     &73.4     &---    &---      &---      &0.14      &9.44       &0.006848 &2.00\\
MIUSCAT               &0.0     &0.0	 &12.4   &0.0       &17.8     &69.7     &---    &---      &---      &0.22      &9.51      &0.02177   &2.34\\
\hline
&&&&&&&NGC 6338\\
\hline
BC03                  &0.00    &0.00     &0.00     &72.3    &0.00     &27. 6    &10.2     &0.0     &89.7     &1.49     &8.21    &0.02373  &1.86\\
M05                   &0.00    &0.00     &0.00     &54.8    &39.4     &5.7      &6.4      &35.2    &58.2     &1.79     &7.82    &0.02323  &2.12\\
C09                   &0.00    &0.00     &0.00     &6.83    &13.3     &79.7     &---      &---     &---      &1.70     &9.65    &0.003544 &2.16\\
BC03io                &0.00    &0.00     &0.00     &0.00    &0.00     &100.0    &0.0      &0.0     &100.0    &1.18     &9.50    &0.0243   &2.23\\
M05io                 &0.00    &0.00     &0.00     &0.00    &45.7     &54.2     &0.0      &22.4    &77.5     &1.56     &9.37    &0.01852  &2.25\\
C09io                 &0.00    &0.00     &0.00     &0.00    &0.00     &100.0    &---      &---     &---      &1.57     &9.96    &0.0004   &2.09\\
MG15                  &1.8     &0.00     &0.00     &0.00    &0.00     &98.1     &---      &---     &---      &0.88     &10.13   &0.005994 &1.76\\
MIUSCAT               &0.7     &0.00     &0.00     &0.00    &18.2     &81.0     &---      &---     &---      &0.97     &9.93    &0.0119   &1.75\\
\hline
&&&&&&&UGC 08234\\
\hline
BC03                  &0.0     &0.0      &0.0    &67.8      &0.0      &32.1     &6.5    &0.0      &93.4     &0.89      &8.22       &0.02292  &1.48\\
M05                   &0.0     &0.0      &0.0    &46.3      &43.0     &10.6     &4.1    &32.2     &63.5     &0.83      &7.95       &0.02678  &1.41\\
C09                   &0.0     &0.0      &0.0    &27.1      &0.0      &72.8     &---     &---     &---      &0.91      &9.06       &0.01242  &1.68\\
BC03io                &0.0     &0.0      &0.0    &0.0       &0.0      &100.0    &0.0    &0.0      &100.0    &0.64      &9.90       &0.01971  &1.74\\
M05io                 &0.0     &0.0      &0.0    &0.0       &47.7     &52.2     &0.0    &21.6     &78.3     &0.62      &9.42       &0.02828  &1.59\\
C09io                 &0.0     &0.0      &0.0    &0.0       &0.0      &100.0    &---    &---      &---      &0.75      &9.89       &0.01307  &1.77\\
MG15                  &0.0     &0.0      &0.0    &0.0       &0.0      &100.0    &---    &---      &---      &0.17      &9.88       &0.01127  &1.94\\
MIUSCAT               &0.0     &0.0	 &0.0    &0.0       &54.4     &45.5     &---    &---      &---      &0.21      &9.61       &0.01744  &2.11\\
\hline
\end{tabular}
\end{table*}

\par

Fits performed with the BC03 library of models display higher contribution from young populations, a mild contribution from old populations, no contribution of intermediate-age populations and a higher Av when compared with the other models. With M05, the sample shows a dominance of intermediate age populations, with significant ammounts of young and old populations. When using C09 library, {\sc starlight} finds a dominance of old stellar populations, with mild contribution from young populations. When using low resolution models that do not include young populations, BC03io and C09io models result in a $\sim$97\% contribution from old populations for all the objects, with the other 3\% due to the FC+BB. When using M05io models, on the other hand, 3 objects (NGC 4636, NGC 5905 and NGC 6081) appear dominated by intermediate age populations, with old populations dominating the rest of the sample. With high spectral resolution libraries of models, a dominance of old populations was found, with higher contributions from FC and BB when compared with lower resolution ones. However, results obtained using MIUSCAT library display a trend toward intermediate-age SSPs whereas results obtained using MG15 library display a trend toward old SSPs. For NGC 6338 and UGC 08234, {\sc starlight} found intermediate-age populations dominating the emission (70\% and 54 \% respectively) when using MIUSCAT library, while no contribution was found when using MG15. For NGC 5905, a contribution of 31\% from an intermediate population was found when using MIUSCAT library, while no contribution was found for this age when using MG15. For a better visualization of these differences, we plotted the average contribution from young, intermediate and old populations plus the summed contributions of the FC and BB on Figure~\ref{fig:percentage}

\par

The stellar synthesis using low spectral resolution libraries of models did not fit properly the 2.3$\mu$m CO bands on NGC 5966, NGC 6146, NGC 6338 and UGC 08234 regardless of the low resolution library of models used. The Na{\sc i} and Ca{\sc i} lines in the K-band and the absorptions in the J and H band were not fitted with any low resolution library. A similar result was reported by \citet{riffel15}, who found that none of the models tested by them \citep[BC03, M05 and][]{M11} accurately reproduces all of the stellar features observed in the spectra. In all cases, the fits improved when young populations were included. When using high resolution libraries, the absorption features in H and K bands were fitted with a much better agreement between models and observations. However, in  the J band, because of the low signal-to-noise ratio, the fits are not as good. Note that this is not a big issue because this band is mostly used for fitting the continuum inclination.

\section{Discussion}
\label{sec:discussion}

\subsection{NIR models with low spectral resolution}
\label{sec:NIRlowR}

Our sample is composed mainly of ETGs (NGC\,4636, NGC\,5966, NGC\,6081 NGC\,6146, NGC 6338 and UGC 08234), which are mainly constituted of old populations \citep[][and references therein]{rickes09}.  Analysing Table~\ref{tab:NIRtable}, it is clear that for libraries of models with low spectral resolution (BC03, M05 and C09), the results are linked to the library used rather than to the galaxy properties themselves. This result is different from the optical result reported by \citet{chen10}. They found that changing the library would result in a change of the percentage contributing light fractions, even though the dominant populations are unaltered. They also compared high and low resolution libraries, and the dominant populations still remained unchanged. This is not the case for the NIR. We found from \cref{fig:4636,fig:5905,fig:5966,fig:6081,fig:6146,fig:6338,fig:08234} (see also Table~\ref{tab:NIRtable}) that for the NIR spectral range, the dominant population is highly dependent on the chosen models.

\par

According to our results, BC03 models give the highest contribution from young populations in the whole sample. The higher young SP contributions are also linked to a higher reddening value, as we can see from Table~\ref{tab:NIRtable}. This result is in agreement with those found by \citet{riffel15}. However, they are in contrast with the ones reported by \citet[e.g.][]{capozzi16}, who found that the inclusion of TP-AGB stars tend to produce results with older ages. This difference might be connected to the fact that they used panchromatic Spectral Energy Distributions, while we used only NIR spectra to make the comparison.

\par

Indeed, for models with low spectral resolution, the large variation in age when young populations are not included implies that the synthesis is unable to distinguish between an old population and a reddened younger one or a more metallic one, since in all cases, the younger populations are followed by a higher extinction or a higher metallicity. This result was first discussed by \citet{worthey94} and seems to still hold in the NIR. 

\par

Regarding the metallicity, when young populations are present in the fitting process, both BC03 and M05 find values close to solar. C09, on the other hand, tends to find subsolar values of metallicity. When young ages are removed, the same results still hold. Concerning the high resolution libraries, MIUSCAT also finds values close to Z$_\odot$ whereas MG15, tends to find subsolar values.

\par

Also, all model sets with low spectral resolution found high ($A_v\gtrsim$1.0mag) values of extinction. These results are uncompatible with literature results who show that ETGs have negligible amounts of dust \citep{PadillaStrauss08}.

\subsection{NIR models with high spectral resolution}
\label{sec:NIRhighR}

For EPS models with high spectral resolution (MG15 and MIUSCAT), the above scenario improved considerably. For three objects (NGC 4636, NGC 5966 and NGC 6081), besides the small (<10\%) contribution from dust, only old SSPs were found using both libraries of models. For NGC 6146, both libraries predict a contribution of intermediate populations of $\sim$18\%. For NGC 5905, the only spiral galaxy in our sample, a contribution of 31\% from intermediate-age stars was obtained with MIUSCAT, but this same percentage was not detected when using MG15. Lastly, for UGC 08234, the fit with MG15 SSPs appear dominated by old populations. However, when MIUSCAT models are employed, the fits become dominated by intermediate-age stellar populations.

\par

Note that it is not possible to fully test libraries with high spectral resolution, because MG15 and MIUSCAT are only composed of SSPs older than 1 Gyr. However, the results we obtained point toward an improvement when using libraries with high resolution. Also, the dust extinction values found with high spectral resolution libraries are much more consistent with literature results that showed that ETGs have very low amounts of dust.

\subsection{Comparison with optical results}
\label{sec:OpticComparison}

\begin{figure*}
        \noindent
	\includegraphics[width=\linewidth]{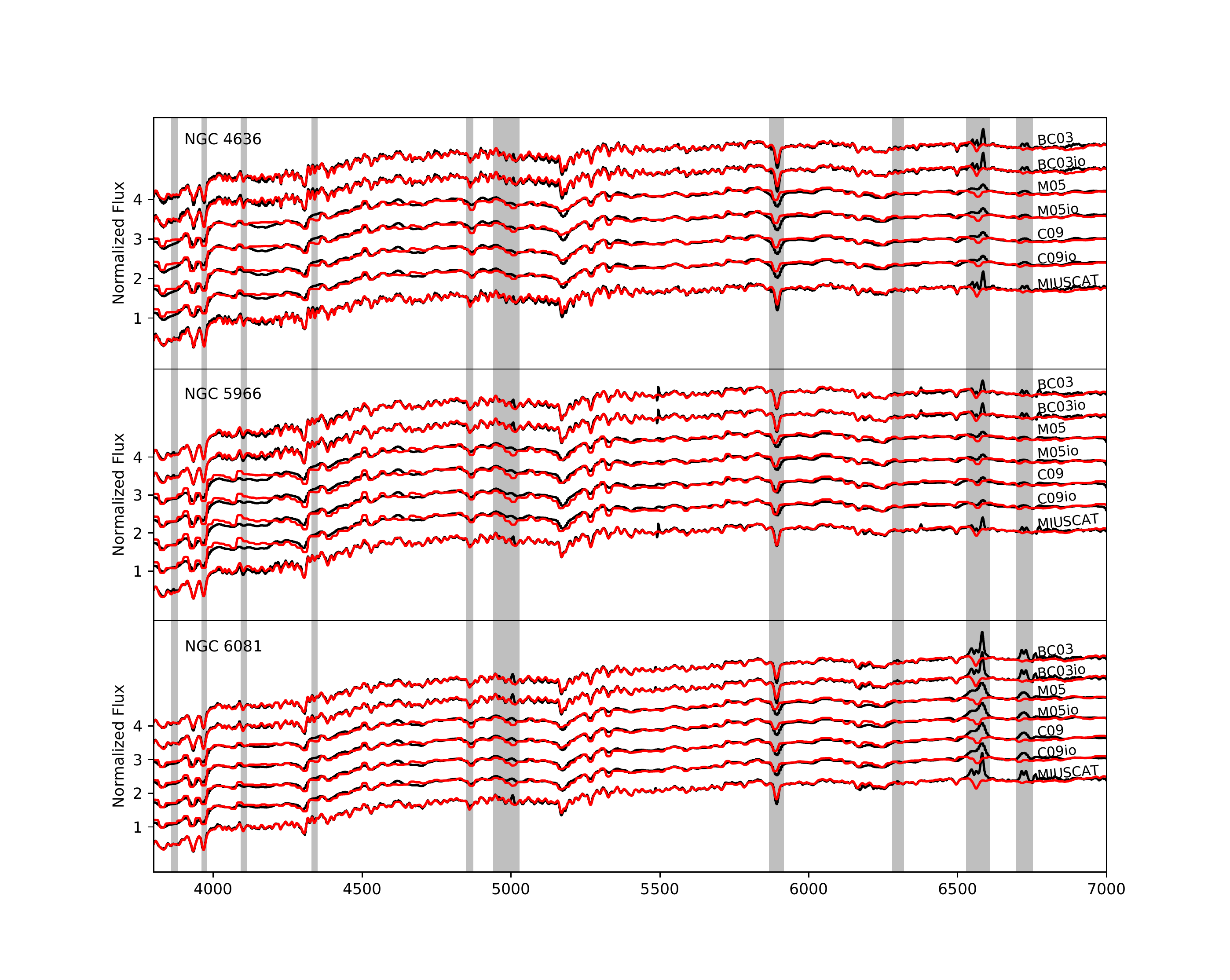}
    \caption{Optical fits for NGC\,4636, NGC\,5966 and NGC\,6081. For each galaxy we present the observed (black) and modeled spectra (red) for the 4 libraries used. }
    \label{fig:OptSpec1}
\end{figure*}

\begin{figure*}
        \noindent
	\includegraphics[width=\linewidth]{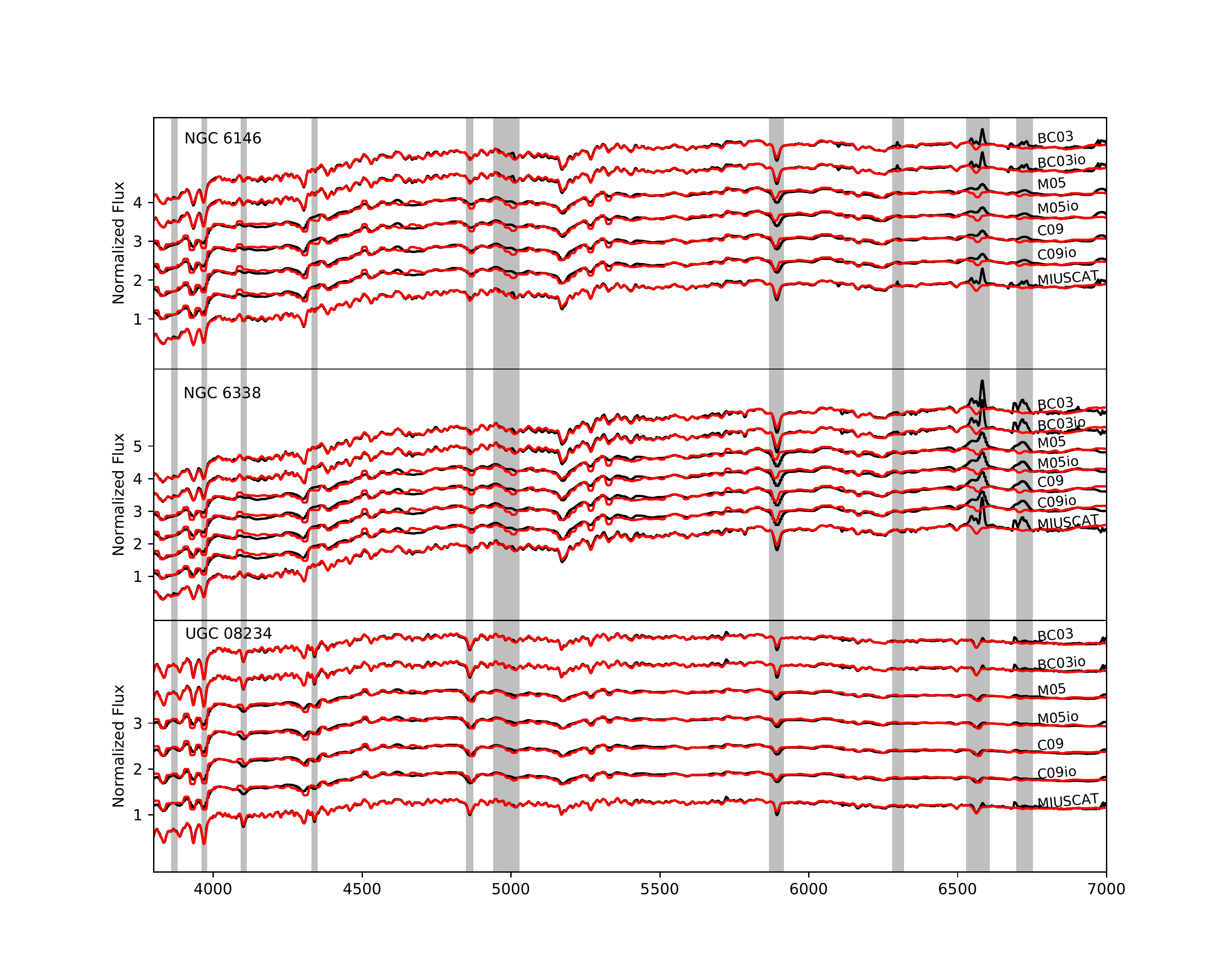}
    \caption{Same of Figure~\ref{fig:OptSpec1} but for NGC\,6146, NGC\,6338 and UGC\,08234.}
    \label{fig:OptSpec2}
\end{figure*}

\begin{figure*}
        \noindent
	\includegraphics[width=\linewidth]{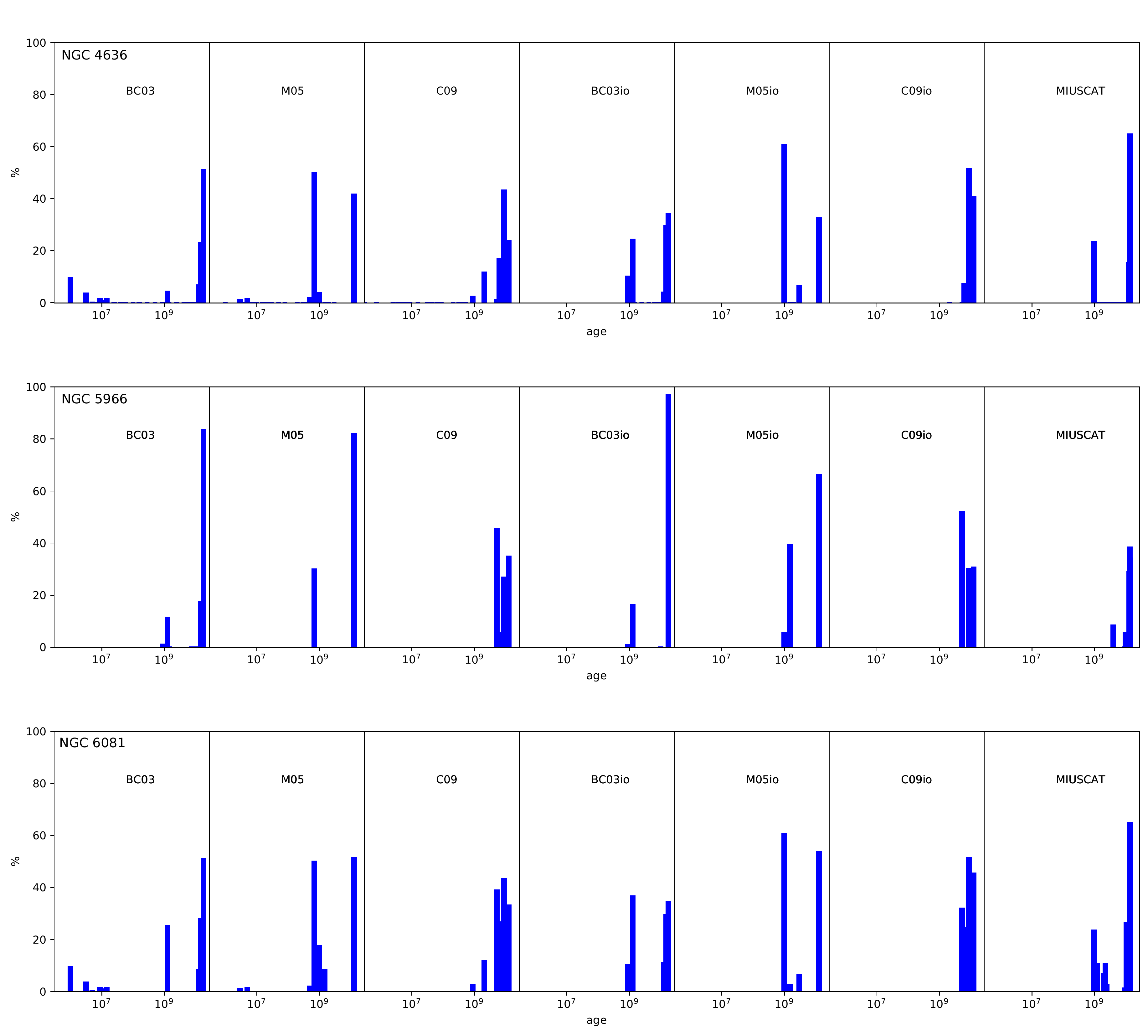}
    \caption{Optical star formation histories for NGC\,4636, NGC\,5966 and NGC\,6081, metallicities summed.}
    \label{fig:opticsfh01}
\end{figure*}

\begin{figure*}
        \noindent
	\includegraphics[width=\linewidth]{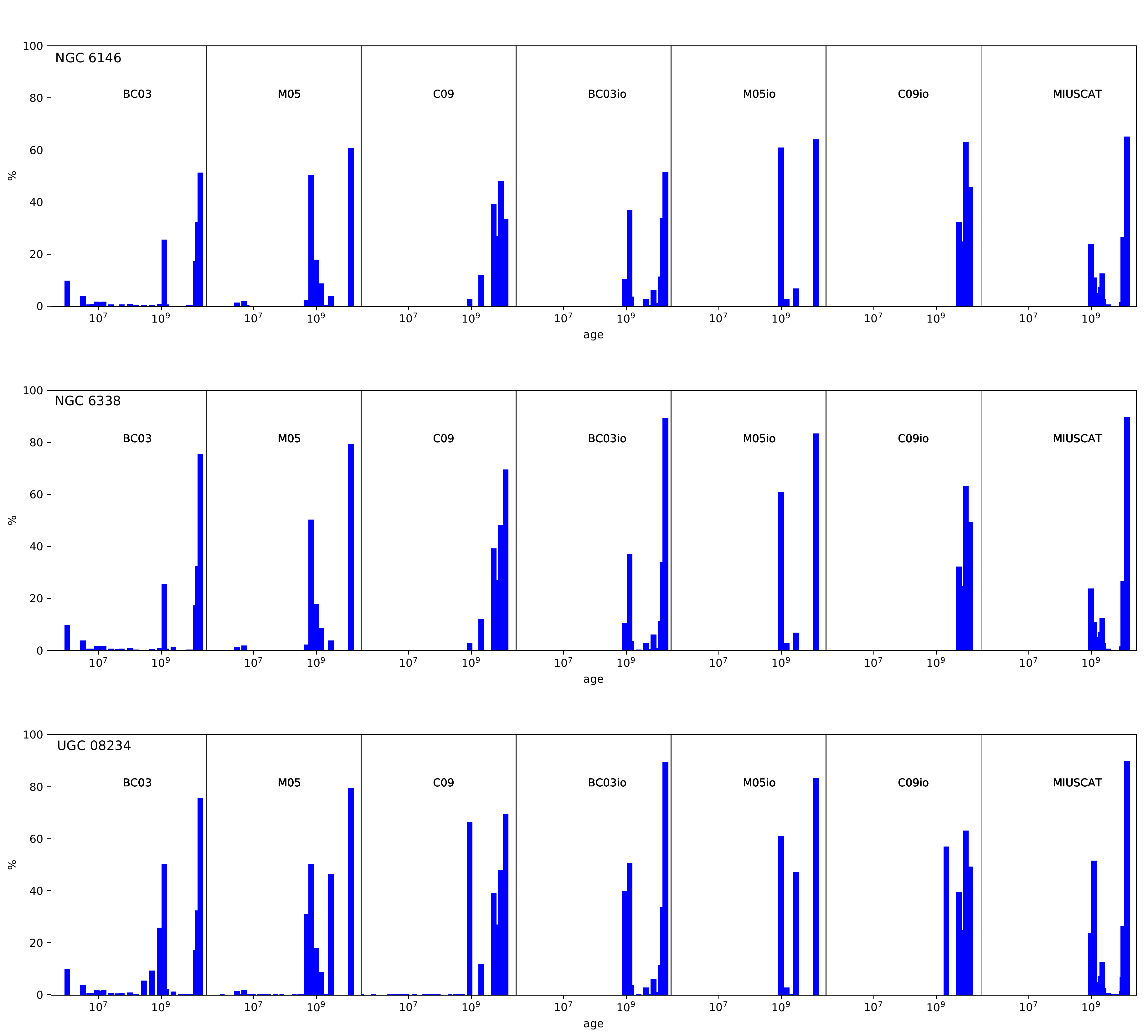}
    \caption{Same as Figure~\ref{fig:opticsfh01} but for NGC\,6146, NGC\,6338 and UGC\,08234.}
    \label{fig:opticsfh02}
\end{figure*}

For the optical region, we also fit the data using the different libraries of models. Since MG15 models are only available in the NIR, we performed the synthesis using only BC03, M05, C09 and MIUSCAT libraries. Also, for a fair comparison between optical and NIR results, we performed the fits using the libraries without the young ages, {\it i.e.} BC03io, M05io and C09io. Figure~\ref{fig:OptSpec1} and Figure~\ref{fig:OptSpec2} present the optical synthesis results for the 6 galaxies with optical spectra. For each galaxy, we show the observed spectra in black and the model spectrum in red. On Figure~\ref{fig:opticsfh01} and Figure~\ref{fig:opticsfh02}, we show the light fraction for each age in blue. The binned light and mass contribution, along with the Av, <t>$_L$, <Z>$_L$ and the Adev for each galaxy are shown in Table~\ref{tab:optictable}.

\par

In the optical synthesis, the different libraries of models produced results that are more similar among each other than those derived from the NIR. The only exception were the fits with M05 and M05io models, that also in the optical found a high fraction of intermediate populations. For the other libraries, in the optical region, the synthesis found a dominant contribution of old populations to the light of NGC\,4636, NGC\,5966, NGC\,6081, NGC\,6146 and NGC\,6338 and a dominance of intermediate-age populations for UGC\,08234. Using M05, on the other hand, the synthesis found similar results for all the objects, with small ($\lesssim$5\%) contributions of young populations and higher fractions of intermediate age (25\%$\lesssim$xi$\lesssim$55\%) and old (40\%$\lesssim$xi$\lesssim$75\%) populations. It is worth mentioning that when using BC03 to fit the optical spectra, a larger contribution from young stellar populations is found compared to the other libraries. However, the dominant old population (or intermediate for UGC\,08234) still remains the same found with the other libraries (see Table~\ref{tab:optictable}), while in the NIR spectral range, the dominant population contributing to the galaxy emission was different (see Table~\ref{tab:NIRtable}). Removing the SSPs with t$<$1Gyr from BC03 did not change the results.

\begin{table*}
\scriptsize
\centering
\caption{Results for the optical synthesis.}
\label{tab:optictable}
\begin{tabular}{lcccccccccr}
\hline
Library	   &xy    &xi        &xo        &my        &mi       &mo       &Av         &$<t>_L$   &$<Z>_L$  &Adev\\
\hline
&&&&&NGC 4636\\
\hline
  BC03     &17.1 &4.3      &78.4     &0.0       &0.1      &99.8     & -0.01     &9.51      &0.02398  &1.93\\
  M05      &2.8  &55.6     &41.4     &0.0       &7.3      &92.6     & 0.40      &9.31      &0.03275  &2.50\\  
  C09      &0.0  &2.4      &97.5     &---       &---      &---      & 0.25      &9.86      &0.01811  &2.37\\  
 BC03io    &0.0  &33.7     &66.2     &0.0       &1.2      &98.8     & 0.02      &9.82      &0.02305  &2.02\\
 M05io     &0.0  &60.7     &39.2     &0.0       &14.2     &85.7     & 0.37      &9.39      &0.0296   &2.59\\
 C09io     &0.0  &5.4      &94.5     &---       &---      &---      & 0.25      &9.86      &0.01895  &2.38\\   
MIUSCATIR  &0.0  &22.6     &77.3     &---       &---      &---      & 0.28      &9.88      &0.02315  &1.95\\
\hline
&&&&&NGC 5966\\
\hline
  BC03     &0.0  &11.0     &88.4     &0.0       &1.0      &98.9     & -0.01     &10.11     &0.02526  &1.84\\
  M05      &0.0  &26.8     &73.2     &0.0       &2.5      &97.4     & 0.26      &9.77      &0.02563  &3.61\\  
  C09      &0.0  &0.0      &100.0    &---       &---      &---      & 0.16      &9.90      &0.02232  &3.89\\    
 BC03io    &0.0  &15.1     &84.8     &0.0       &1.4      &98.5     & -0.00     &10.08     &0.0265   &1.83\\
 M05io     &0.0  &40.5     &59.5     &0.0       &9.9      &90.0     & 0.23      &9.73      &0.02708  &3.65\\
 C09io     &0.0  &0.0      &100.0    &---       &---      &---      &0.16       &9.88      &0.02289  &3.89\\   
MIUSCATIR  &0.0  &0.0      &100.0    &---       &---      &---      & 0.18      &10.09     &0.02663  &1.74\\
\hline
&&&&&NGC 6081\\
\hline
  BC03     &5.0  &23.9     &71.0     &0.0       &1.1      &98.8     & 0.69      &9.75      &0.01817  &1.60\\
  M05      &0.0  &49.8     &50.1     &0.0       &6.5      &93.4     & 0.91      &9.53      &0.02232  &2.48\\  
  C09      &0.0  &0.0      &100.0    &---       &---      &---      & 0.85      &9.87      &0.01503  &2.60\\  
 BC03io    &0.0  &34.6     &65.3     &0.0       &1.6      &98.3     & 0.71      &9.82      &0.01778  &1.69\\
 M05io     &0.0  &47.7     &52.2     &0.0       &6.1      &93.8     & 0.88      &9.59      &0.0181   &2.51\\
 C09io     &0.0  &0.0      &100.0    &---       &---      &---      &0.84       &9.92      &0.01447  &2.59\\
MIUSCATIR  &0.0  &16.7     &83.2     &---       &---      &---      & 0.92      &9.83      &0.02177  &1.49\\
\hline
&&&&&NGC 6146\\
\hline
  BC03     &6.6  &9.6      &83.0     &0.0       &0.3      &99.6     & -0.03     &9.84      &0.0213   &1.29\\
  M05      &0.0  &38.6     &61.4     &0.0       &4.4      &95.5     & 0.28      &9.60      &0.02732  &2.65\\  
  C09      &0.0  &0.0      &100.0    &---       &---      &---      & 0.16      &9.89      &0.02022  &2.72\\  
 BC03io    &0.0  &13.3     &86.6     &0.0       &0.5      &99.5     & -0.02     &10.04     &0.01982  &1.33\\
 M05io     &0.0  &33.3     &66.6     &0.0       &4.5      &95.4     & 0.26      &9.71      &0.02072  &2.70\\
 C09io     &0.0  &0.0      &100.0    &---       &---      &---      &0.16       &9.88      &0.02289  &3.89\\   
MIUSCATIR  &0.0  &13.0     &86.9     &---       &---      &---      & 0.22      &9.91      &0.02246  &1.13\\
\hline
&&&&&NGC 6338\\
\hline
  BC03     &1.3  &1.0      &94.0      &0.0       &0.0      &99.9     & 0.33      &10.16     &0.02569  &1.77\\
  M05      &0.0  &25.2     &74.8      &0.0       &2.3      &97.6     & 0.67      &9.79      &0.02561  &3.17\\  
  C09      &0.0  &0.0      &100.0     &---       &---      &---      & 0.53      &10.05     &0.01783  &3.36\\  
 BC03io    &0.0  &0.9      &99.0      &0.0       &0.0      &99.9     & 0.33      &10.22     &0.02434  &1.78\\
 M05io     &0.0  &21.5     &78.4      &0.0       &2.0      &97.9     & 0.65      &9.87      &0.02038  &3.22\\
 C09io     &0.0  &0.0      &100.0     &---       &---      &---      &0.53       &10.03     &0.01938  &3.34\\   
MIUSCATIR  &0.0  &3.7      &96.2      &---       &---      &---      & 0.59      &10.10     &0.02628  &1.63\\
\hline
&&&&&UGC 08234\\
\hline
  BC03     &0.0  &86.5     &13.4      &0.0        &26.2     &73.7     & -0.07     &9.15      &0.04017  &1.18\\
  M05      &0.0  &40.0     &60.0      &0.0        &8.5      &91.4     & 0.03      &9.28      &0.02446  &2.12\\  
  C09      &0.0  &65.5     &34.4      &---        &---      &---      & 0.23      &9.28      &0.02118  &2.32\\  
 BC03io    &0.0  &85.0     &14.9      &0.0        &25.9     &74.1     & -0.07     &9.21      &0.0368   &1.18\\
 M05io     &0.0  &42.7     &57.2      &0.0        &14.0     &85.9     & -0.05     &9.34      &0.01896  &2.20\\
 C09io     &0.0  &61.5     &38.4      &---        &---      &---      & 0.21      &9.29      &0.01998  &2.32\\   
MIUSCATIR  &0.0  &67.1     &32.8      &---        &---      &---      & 0.07      &9.40      &0.02134  &1.00\\
\hline
\end{tabular}
\end{table*}

\begin{figure*}
        \noindent
	\includegraphics[width=\linewidth]{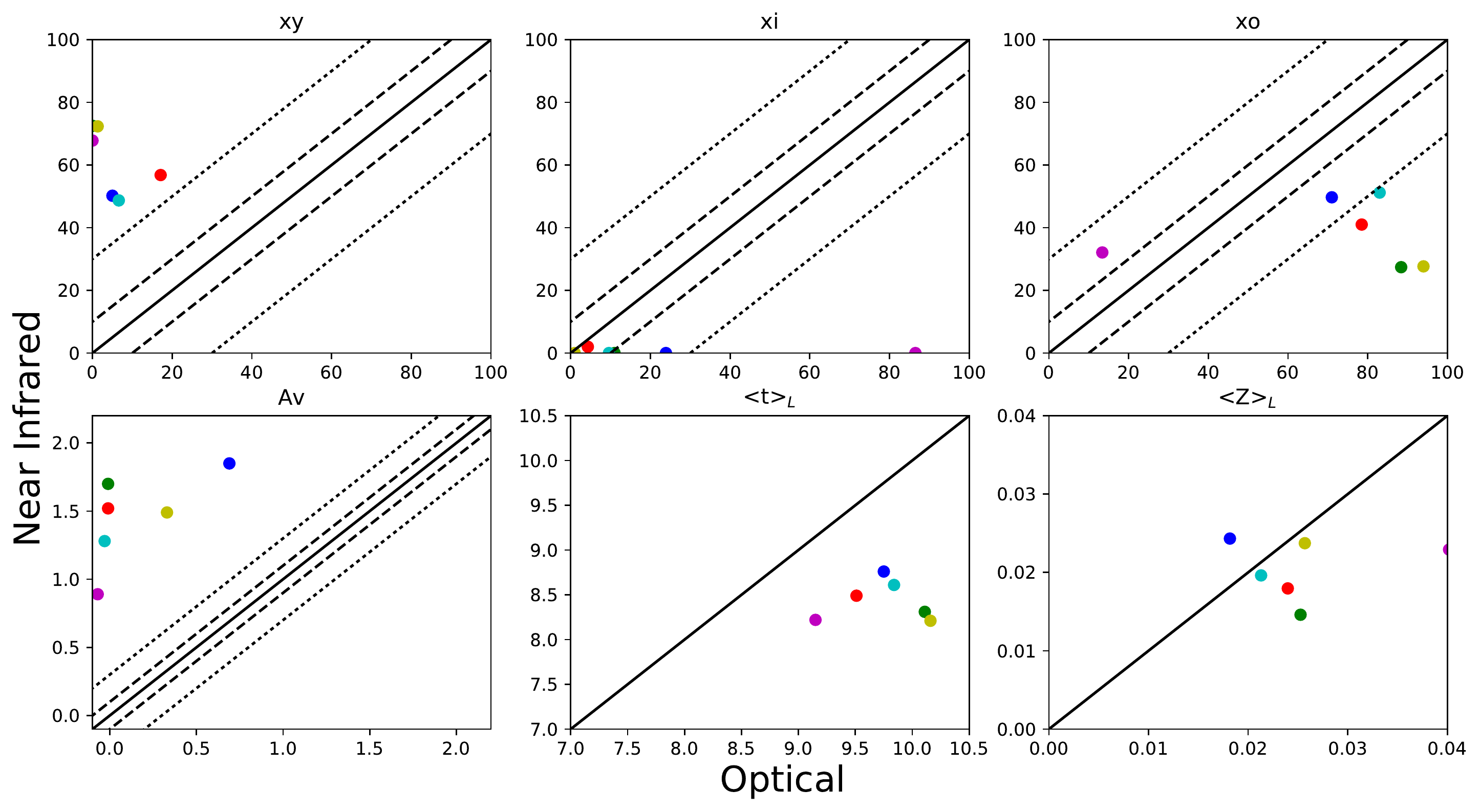}
    \caption{Comparison between optical and NIR results found using BC03 library of models. In the upper panels, we present from left to right the xy, xi and xo results. In the bottom panels, we show the {\rm A$_V$}, $<t>_L$  and $<Z>_L$ results. Over all panels, we plotted a solid line, representing a perfect correlation between optical and NIR. Also, we plotted dashed and dotted lines representing error margins of 10 and 30\% in the xy, xi and xo panel and error margins of 0.1 and 0.3 magnitudes in the {\rm A$_V$} panel. Each galaxy was plotted in a different color: NGC\,4636 - red, NGC\,5966 - green, NGC\,6081 - blue, NGC\,6146 - cyan, NGC\,6338 - yellow, UGC\,08234 - magenta.}
    \label{fig:BC03}
\end{figure*}

\begin{figure*}
        \noindent
	\includegraphics[width=\linewidth]{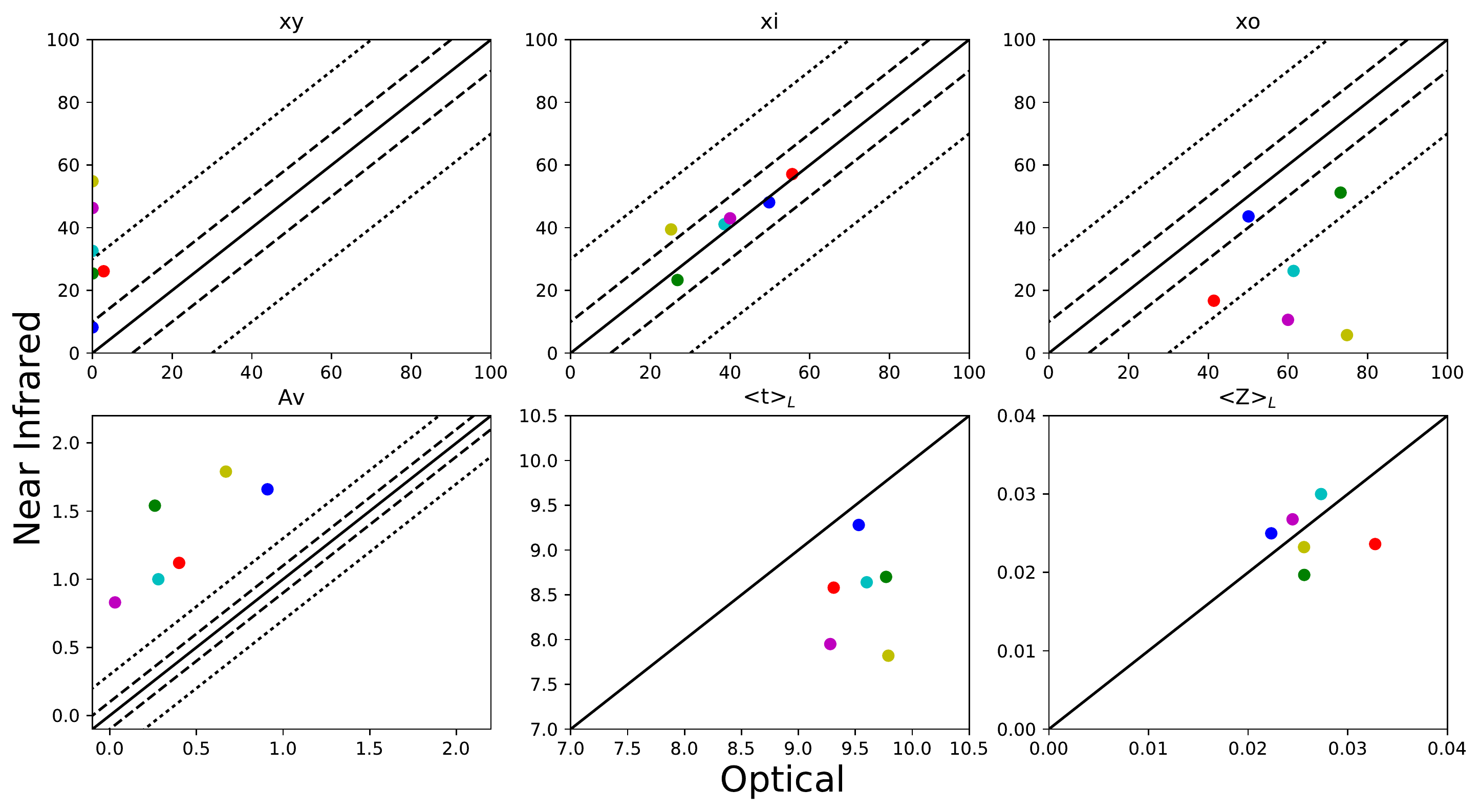}
    \caption{Same of Figure~\ref{fig:BC03}, but for M05 library of models.}
    \label{fig:M05}
\end{figure*}

\begin{figure*}
        \noindent
	\includegraphics[width=\linewidth]{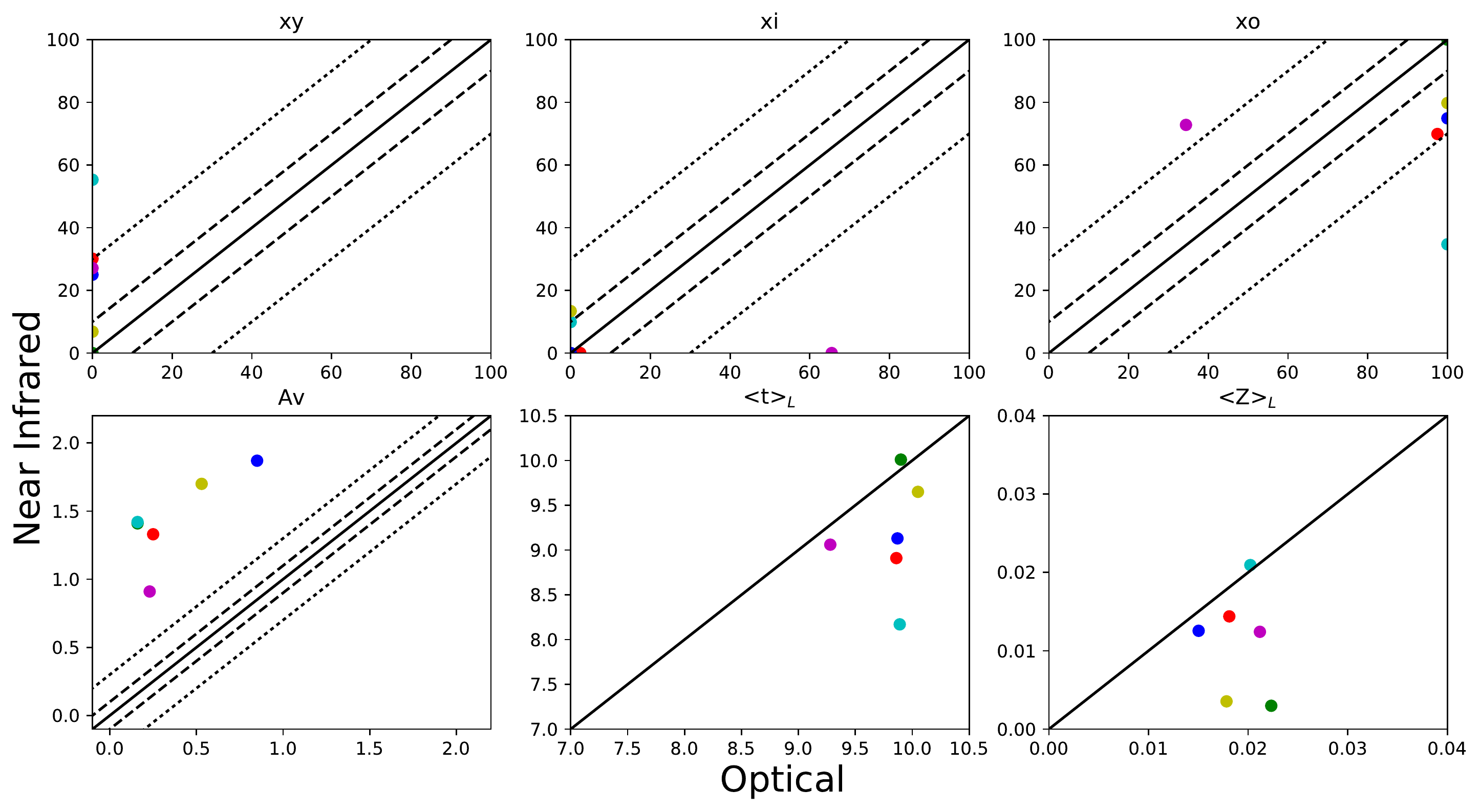}
    \caption{Same of Figure~\ref{fig:BC03}, but for C09 library of models.}
    \label{fig:C09}
\end{figure*}

\begin{figure*}
        \noindent
	\includegraphics[width=\linewidth]{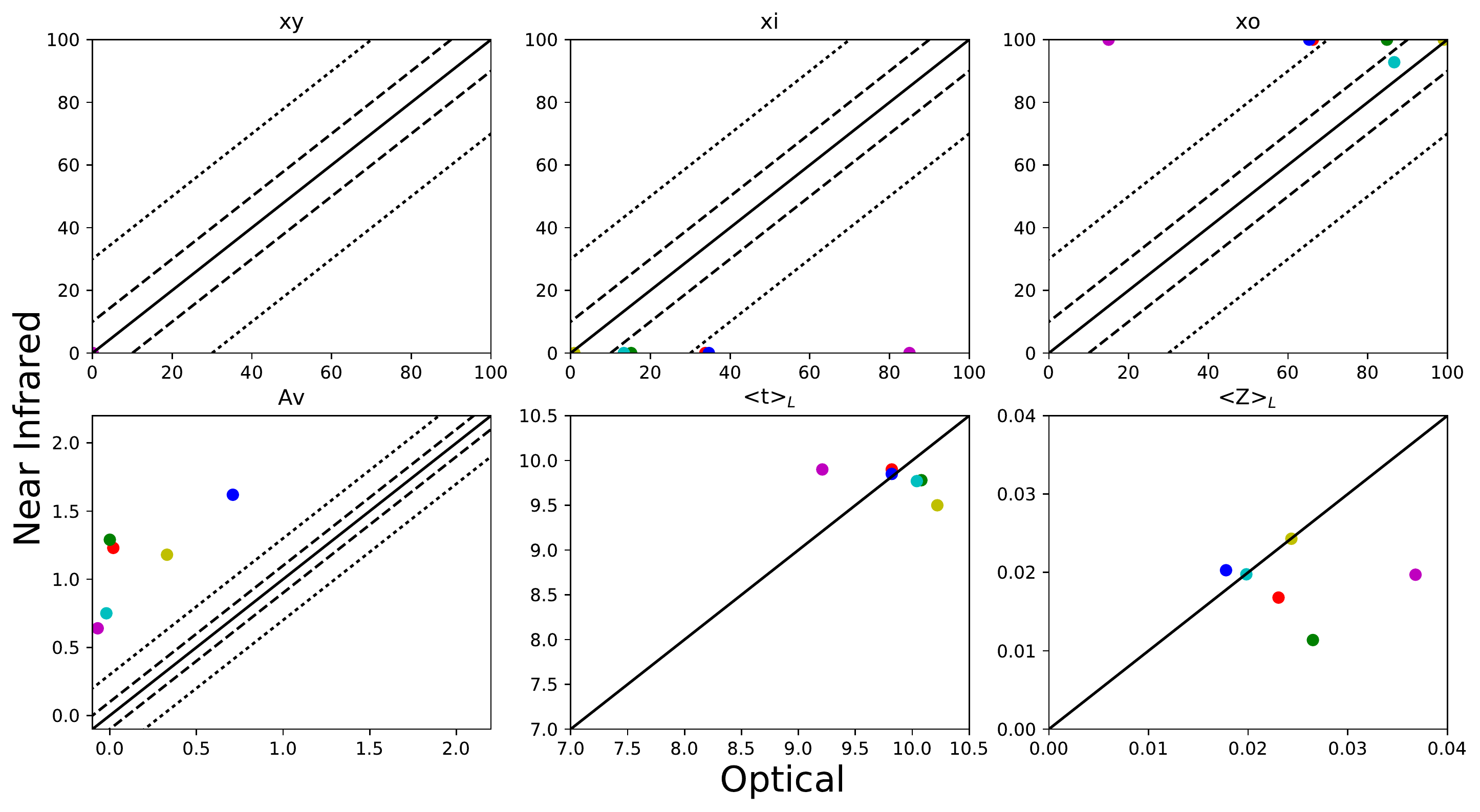}
    \caption{Same of Figure~\ref{fig:BC03}, but for BC03io library of models.}
    \label{fig:BC03io}
\end{figure*}

\begin{figure*}
        \noindent
	\includegraphics[width=\linewidth]{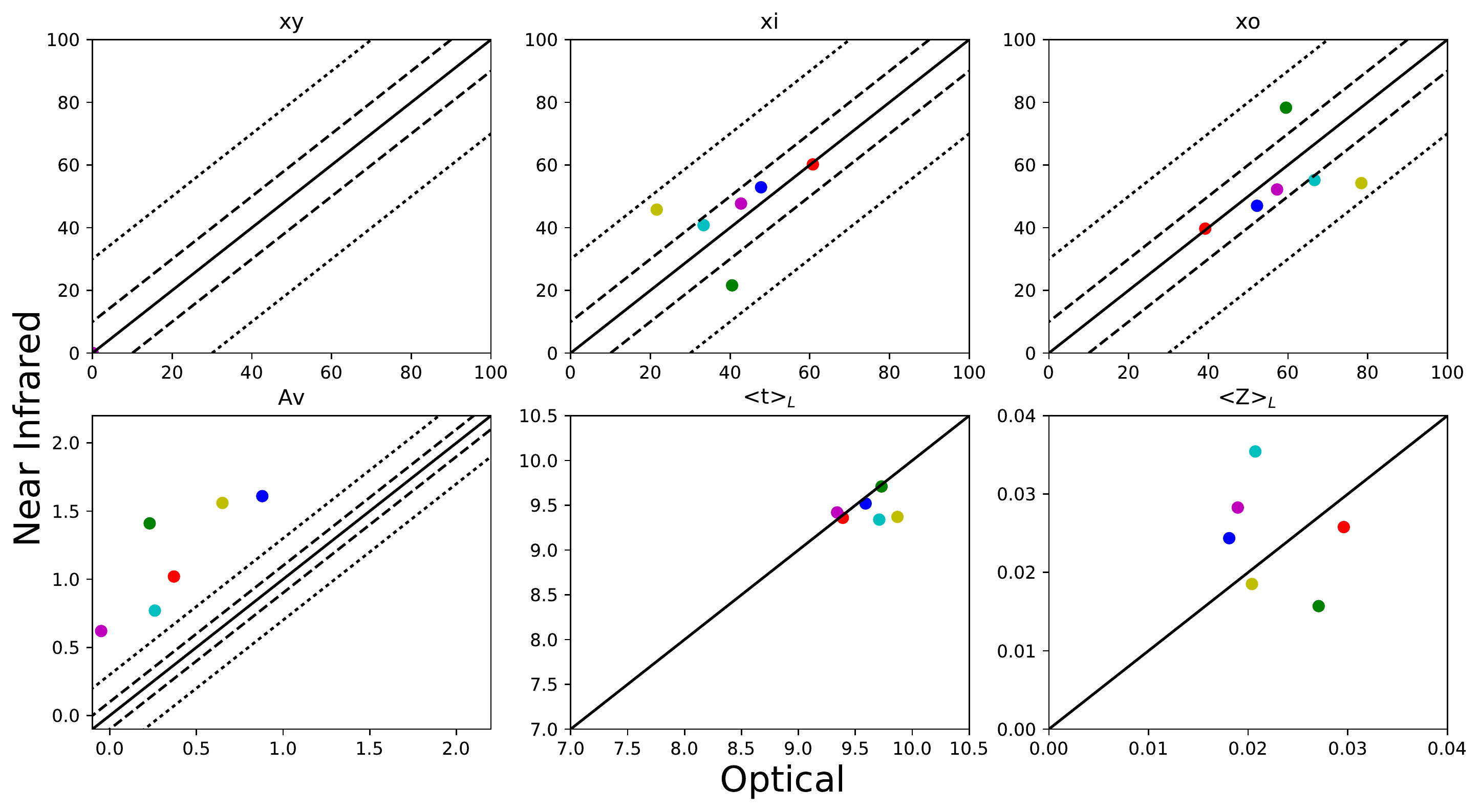}
    \caption{Same of Figure~\ref{fig:BC03}, but for M05io library of models.}
    \label{fig:M05io}
\end{figure*}

\begin{figure*}
        \noindent
	\includegraphics[width=\linewidth]{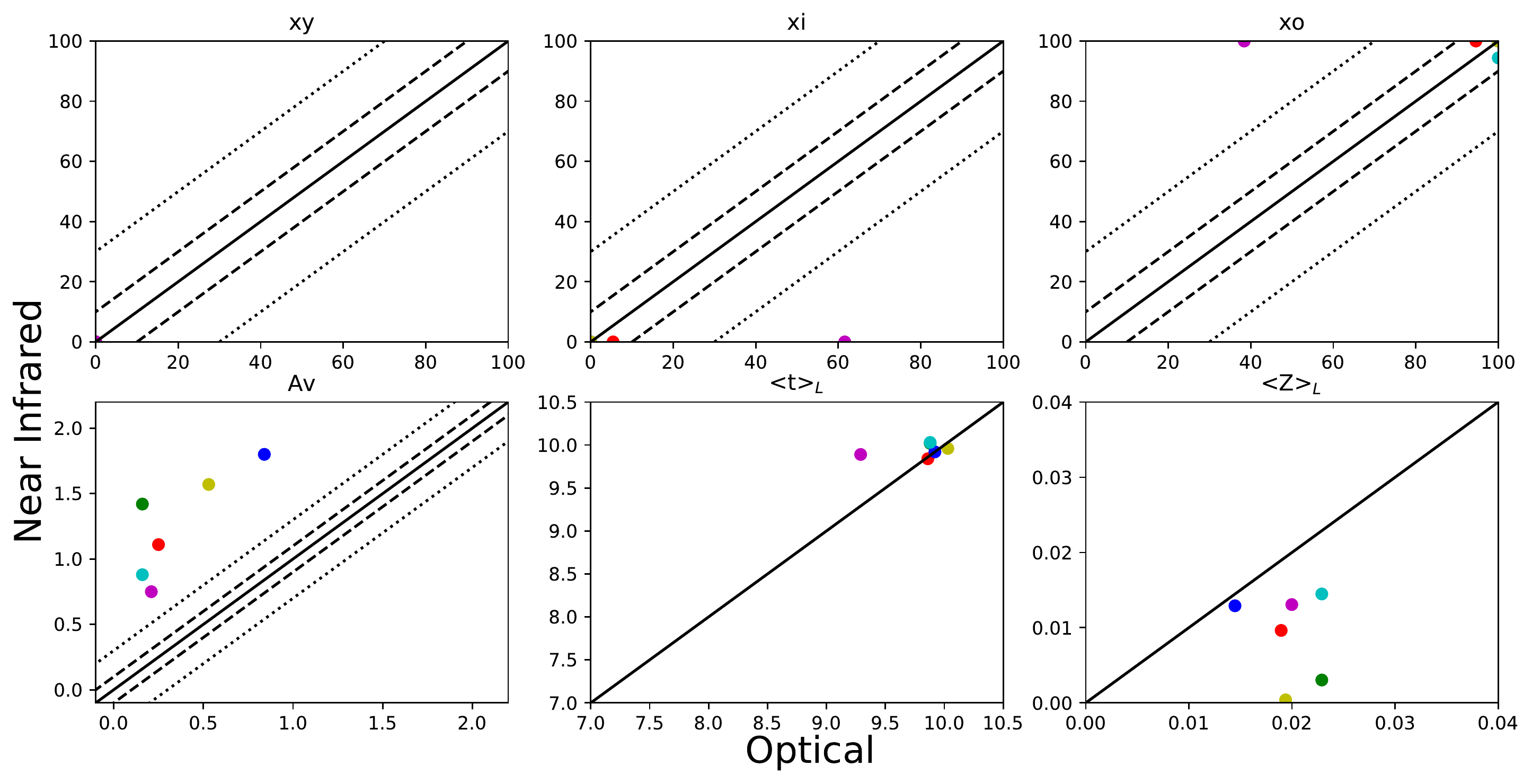}
    \caption{Same of Figure~\ref{fig:BC03}, but for C09io library of models.}
    \label{fig:C09io}
\end{figure*}

\begin{figure*}
        \noindent
	\includegraphics[width=\linewidth]{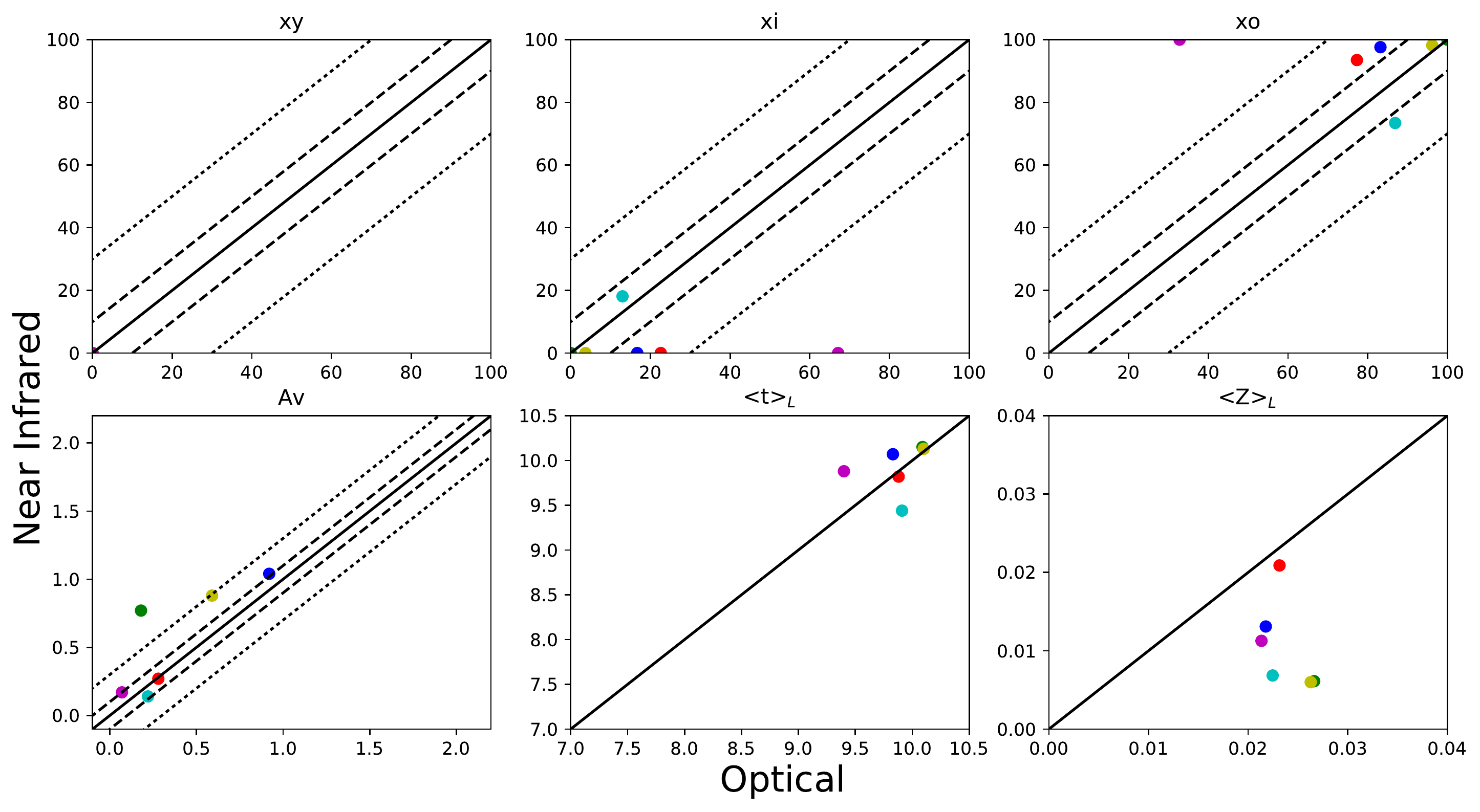}
    \caption{Same of Figure~\ref{fig:BC03}, but for MG15 library of models.}
    \label{fig:MG15}
\end{figure*}

\begin{figure*}
        \noindent
	\includegraphics[width=\linewidth]{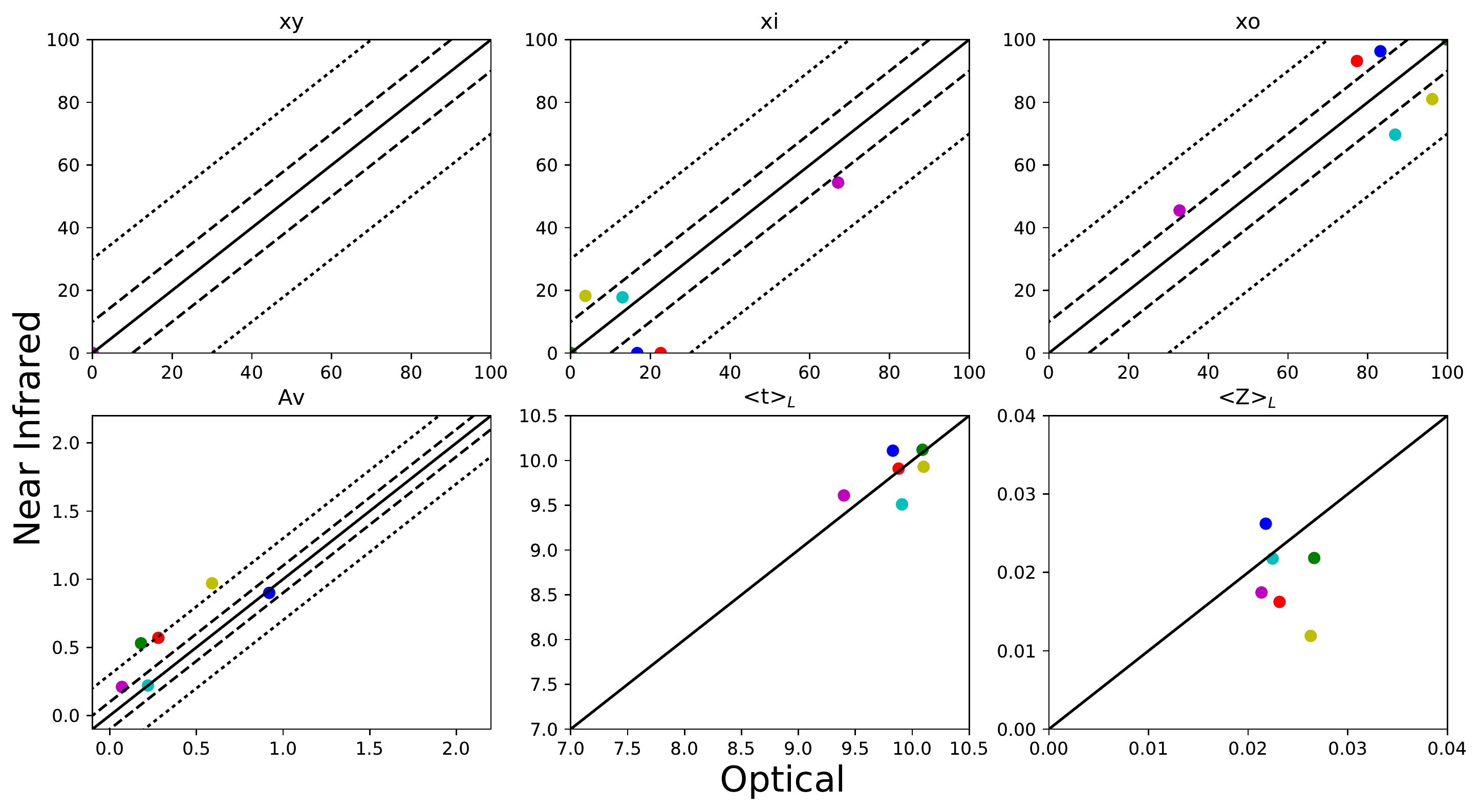}
    \caption{Same of Figure~\ref{fig:BC03}, but for MIUSCAT library of models.}
    \label{fig:MIUSCAT}
\end{figure*}

For a proper comparison between optical and NIR results, we present in \cref{fig:BC03,fig:M05,fig:C09,fig:BC03io,fig:M05io,fig:C09io,fig:MG15,fig:MIUSCAT} the values of xy, xi, xo, {\rm A$_V$}, $<t>_L$  and $<Z>_L$ found by {\sc starlight} both for the optical and NIR ranges. The x axis displays the optical results and the y axis displays the NIR results found by {\sc starlight} for the 8 different libraries of models used. For MG15 library, since it only has the NIR range of the spectra, we compared it to the optical results obtained with MIUSCAT. Over the 6 panels of each picture, we plotted a solid line, representing a perfect correlation between optical and NIR. Over the xy, xi and xo panels, we also plotted two dashed lines showing the 10\% error margin and two dotted lines showing the 30\% error margin. Over the {\rm A$_V$} panel, we plotted two dashed lines showing the 0.1 magnitude error margin and two dotted lines showing the 0.3 magnitude error margin. For the $<t>_L$  and $<Z>_L$ panels, since they are logarithmic scales, we did not display error margins.

From \cref{fig:BC03,fig:M05,fig:C09,fig:BC03io,fig:M05io,fig:C09io,fig:MG15,fig:MIUSCAT}, it is possible to see that libraries with low spectral resolution and with young populations usually disagrees in the optical and the NIR about the fractions of each stellar populations that contributes to the galaxy integrated light. It is clear that the fits using BC03, BC03io, C09 and C09io overestimate the amount of younger stellar populations in the NIR when compared to the fits in the optical. On the other hand, the fits with M05 and M05io are more self consistent when comparing optical and NIR results, where in both cases sizable amounts of intermediate age stars are found.  All the tests with the lower spectral resolution models displayed a tendency to find extinctions $\sim$1 magnitude higher when fitting the NIR spectral range, compared to optical one. Regarding the metallicity, fits with C09 find lower metallicities in the NIR compared to the optical, whereas BC03 and M05 fits using NIR or optical data converge to nearly the same metallicities.

\par

When SSPs with t$\leq$1Gyr were removed from the libraries, the stellar populations found in the optical and NIR were much closer to each other. From the three libraries with low resolution and without young SSPs, M05io was the one that found the most consistent results between the two wavelength ranges, with all the points in the xi and xo panels inside the 30\% error margin. However, since the amounts of intermediate-age SSPs found with this library are close to $\sim$50\% and the sample is mainly composed of ETGs, these results suggest that both wavelength ranges tend equally to overestimate the amount of intermediate-age SSPs.

\par

For BC03io, we found larger amounts of intermediate age stars in the optical than in the NIR, where only old SSPs contributed to the fits. For the tests with C09io, we found a dominant contribution from old SSPs in both spectral ranges. The only exception was UGC\,08234, that a contribution of 85\% of intermediate-age SSPs was required in the optical but only old SSPs where used in the NIR. When dealing with the reddening, these libraries also tend to overestimate the value of {\rm A$_V$} by $\sim$1 magnitude.
\par
The high resolution models, on the other hand, perform considerably better, producing self consistent fits from NIR and optical fits. The reddening found with MG15 models is almost the same for the two wavelength ranges for the whole sample, while MIUSCAT fits required an {\rm A$_V$} 0.3mag larger for two objects. From the two last panels of \cref{fig:MG15,fig:MIUSCAT}, it is possible to see that for both libraries, optical and NIR tend to find similar values of $<t>_L$ and MG15 tends to find lower values of $<Z>_L$ in the NIR when compared to the optical.
\par
Unfortunately, there are no libraries of high resolution models that include SSPs younger than 1Gyr, meaning that the full scenario cannot be tested. However, our results suggest that high resolution models are essencial to correctly disentangle the stellar population of a galaxy in the NIR, allowing in addition reliable reddening values. Otherwise, it is not possible to fit the absorptions and too much weight is given to the featureless continuum.

\par 

\citet{baldwin+17} found that the impact of age variation in the near-infrared is largely dependent on the shape of the continuum.  Our results show that the scenario is more complex, where the shape of the continuum is dependent on a combination of age and extinction, while the depth, shape and width of the absorption features are crucially shaped by the age of the population. For low resolution SSPs these features will be diluted. As a result, {\sc starlight} will have more difficulties to differentiate among the individual SSPs. Therefore, we reinforce the need of adequate NIR stellar libraries for hotter stars, allowing thus the production of models with ages younger than 1\,Gyr. This is specially important to properly fit the stellar content of active starforming galaxies.

\subsection{Absorption Band Measurements}
\label{sec:AbsorptionBands}

For comparison with future NIR SP studies, we computed the equivalent widths (W$_\lambda$) of the NIR absorption features measured with a python version of pacce code \citep{pacce}. The values of W$_\lambda$ are presented on Table~\ref{tab:NIREW}. They were calculated based on the line limits and continuum bandpasses of Table~\ref{tab:ewdefs}. Mild correlations were found between CO16a and MgI, CO16b and SII, CO16c and NaI, CO16d and NaI, FeI and CaI, and the CO22 bands. Since the sample is small, these correlations should be seen with caution.

\begin{table*}
\centering
\caption{Equivalent widths for the NIR absorption bands measured in \AA.}
\label{tab:NIREW}
\begin{tabular}{lcccccccccccr}
\hline
Galaxy/Line & CO16a     &   CO16b    &    SiI     &    CO16c   &    CO16d   &    FeI     &   MgI    &    NaI     &    CaI     &    CO22a        &    CO22b    &    CO22c\\
Average Error & 0.10    &   0.12     &    0.13    &    0.32    &    0.22    &    0.12    &   0.8    &    0.21    &    0.21    &    0.86         &    1.24     &    1.58 \\
\hline
NGC4636    &    2.06    &    3.08    &    1.97    &    6.41    &    4.71    &    0.42    &   2.47   &    4.55    &    2.86    &    16.42	&    18.81    &    18.01\\
NGC5905    &    1.69    &    3.69    &    3.03    &    7.61    &    5.03    &    0.47    &   2.03   &    5.79    &    4.08    &    17.50	&    6.78     &    8.94 \\
NGC5966    &    1.74    &    4.29    &    3.72    &    7.17    &    3.89    &    0.37    &   2.41   &    4.68    &    2.23    &    7.79         &    4.73     &    6.68 \\
NGC6081    &    0.23    &    3.18    &    2.77    &    6.53    &    3.40    &    0.53    &   1.95   &    3.68    &    3.26    &    19.77	&    24.23    &    27.59\\
NGC6146    &    2.38    &    4.23    &    4.07    &    5.89    &    3.72    &    0.39    &   2.55   &    2.50    &    1.65    &    34.72	&    46.77    &    50.05\\
NGC6338    &    2.31    &    3.13    &    2.90    &    5.53    &    4.34    &    0.47    &   2.10   &    3.57    &    3.76    &    13.79	&    13.60    &    8.64 \\
UGC08234   &    1.29    &    3.36    &    3.36    &    5.24    &    3.91    &     --     &   1.82   &    2.20    &    3.69    &    17.77	&    25.97    &    35.33\\
\hline
\end{tabular}
\end{table*}

\begin{table*}
  \centering
  \setlength{\tabcolsep}{2pt}
  \caption{Line limits and continuum bandpasses. \label{tab:ewdefs}}
  \begin{tabular}{cccc}
  \hline
  \hline
  \noalign{\smallskip}
  Centre &     Main        & line limits   & continuum bandpass \\
   (\AA) &   Absorber    & (\AA)         &  (\AA) \\
  \hline
  \noalign{\smallskip}            
  15587 &    CO                     & 15555-15620 & 15110-15170, 15390-15410,16270-16310,16570-16580\\
  15772 &    CO                     & 15735-15810 & 15110-15170, 15390-15410,16270-16310,16570-16580\\
  15890 &    Si{\sc i}              & 15850-15930 & 15110-15170, 15390-15410,16270-16310,16570-16580\\
  16215 &    CO                     & 16145-16285 & 15110-15170, 15390-15410,16270-16310,16570-16580\\
  16385 &    CO                     & 16340-16430 & 15110-15170, 15390-15410,16270-16310,16570-16580\\
  17054 &    Fe{\sc i}              & 17025-17083 & 16970-17083, 17140-17200\\
  17106 &    Mg{\sc i}              & 17083-17130 & 16970-17083, 17140-17200\\
  22025 &    Na{\sc i}              & 21950-22100 &  21700-21930,22150-22200\\
  22620 &    Ca{\sc i}              & 22570-22670 &  22450-22550,22680-22730\\
  23015 &    CO                     & 22870-23160 & 22690-22790, 23655-23680,23890-23920\\
  23290 &    CO                     & 23160-23420 & 22690-22790, 23655-23680,23890-23920\\
  23535 &    CO                     & 23420-23650 & 22690-22790, 23655-23680,23890-23920\\
  \hline
  \end{tabular}
  \begin{list}{Table Notes:}
  \item The NIR indexes are based on \citet{riffel08, riffel09,riffel11,riffel15}.
  \end{list}
\end{table*}

\section{Final remarks}
\label{sec:finalremarks}

In this work we compared the stellar population of a sample of 6 ETGs and one spiral galaxy, both in the optical and NIR spectral ranges. We chose for the NIR 8 different bases of SSPs with different spectral resolutions and using different sets of isochrones to separately discuss their effects on the synthesis results. For 4 of the ETGs, we performed spectral synthesis in the optical using SDSS spectra and compared the results with those of the NIR. The approach followed here is based on the {\sc starlight} code, which considers the whole observed spectrum, both the continuum and absorption features.

\par

The main results can be sumarized as follows: for spectral synthesis using bases with low spectral resolution, the results are more linked to the library of models used rather than than to the object properties themselves. While BC03 models display a trend toward higher contributions of young populations, M05 usually finds more contributions from intermediate age populations whereas C09 tend to find higher contributions from old stellar populations. On all the cases, the values of Adev are compatible, meaning that none of these bases offer a more reliable result. When using bases with high spectral resolution, MG15 and MIUSCAT produced more consistent results if compared to low spectral resolution libraries. Out of the 7 galaxies, the two libraries fitted consistent results for six of them.

\par

The optical synthesis for the elliptical galaxies revealed a dominance of old stellar populations. The only exception was M05 library, which still found a high fraction of intermediate age populations. This may indicate that the TP-AGB treatment plays an important role, even in the optical region.

\par

When comparing optical and NIR results, we found that NIR fits using low spectral resolution libraries tend to overestimate the amount of young SSPs and the reddening. The only exception were the M05 models, wich produced self consistent fits, but predicted sizable ammounts of intermediate age stars for the galaxies. For libraries with high spectral resolution, since they do not include young SSPs, this scenario cannot be fully tested. However, the reddening found was compatible with literature. Also, high spectral resolution libraries produced results much more consistent if compared to models with low spectral resolution.

\par

We tabulated the equivalent widths (W$_\lambda$) of the NIR absorption features and the optical emission line fluxes to be used in future studies. Lastly, from the emission line ratios, we classified, for the first time in the literature, NGC 6081 and NGC 6338 as LINERs.

\section*{Acknowledgements}

LGDH thanks CAPES and CNPq. RR thanks CNPq and FAPERGS for partial funding this project. ARA thanks CNPq for partial support to this work. LPM thanks CNPQ and FAPESP for partial funding of this project. CK acknowledges support through the research project AYA2017-79724-C4-4-P from the Spanish PNAYA. This research made use of the NASA/IPAC Extragalactic Database (NED), which is operated by the Jet Propulsion Laboratory,  California Institute of Technology, under contract with the National Aeronautics and Space Administration. This study uses data provided by the Calar Alto Legacy Integral Field Area (CALIFA) survey (http://califa.caha.es/). Based on observations collected at the Centro Astron\'omico Hispano Alem\'an (CAHA) at Calar Alto, operated jointly by the Max-Planck-Institut f\"ur Astronomie and the Instituto de Astrof\'isica de Andaluc\'ia (CSIC). We thank the referee Reynier Peletier for carefully reading our paper and for giving such constructive comments which substantially helped improving the quality of the paper.




\bibliographystyle{mnras}
\bibliography{luisgdh} 

\begin{thebibliography}{}
\makeatletter
\relax
\def\mn@urlcharsother{\let\do\@makeother \do\$\do\&\do\#\do\^\do\_\do\%\do\~}
\def\mn@doi{\begingroup\mn@urlcharsother \@ifnextchar [ {\mn@doi@}
  {\mn@doi@[]}}
\def\mn@doi@[#1]#2{\def\@tempa{#1}\ifx\@tempa\@empty \href
  {http://dx.doi.org/#2} {doi:#2}\else \href {http://dx.doi.org/#2} {#1}\fi
  \endgroup}
\def\mn@eprint#1#2{\mn@eprint@#1:#2::\@nil}
\def\mn@eprint@arXiv#1{\href {http://arxiv.org/abs/#1} {{\tt arXiv:#1}}}
\def\mn@eprint@dblp#1{\href {http://dblp.uni-trier.de/rec/bibtex/#1.xml}
  {dblp:#1}}
\def\mn@eprint@#1:#2:#3:#4\@nil{\def\@tempa {#1}\def\@tempb {#2}\def\@tempc
  {#3}\ifx \@tempc \@empty \let \@tempc \@tempb \let \@tempb \@tempa \fi \ifx
  \@tempb \@empty \def\@tempb {arXiv}\fi \@ifundefined
  {mn@eprint@\@tempb}{\@tempb:\@tempc}{\expandafter \expandafter \csname
  mn@eprint@\@tempb\endcsname \expandafter{\@tempc}}}

\bibitem[\protect\citeauthoryear{{Baldwin}, {Phillips}  \&
  {Terlevich}}{{Baldwin} et~al.}{1981}]{bpt}
{Baldwin} J.~A.,  {Phillips} M.~M.,   {Terlevich} R.,  1981, \mn@doi [\pasp]
  {10.1086/130766}, \href {http://adsabs.harvard.edu/abs/1981PASP...93....5B}
  {93, 5}

\bibitem[\protect\citeauthoryear{{Baldwin}, {McDermid}, {Kuntschner},
  {Maraston}  \& {Conroy}}{{Baldwin} et~al.}{2017}]{baldwin+17}
{Baldwin} C.~M.,  {McDermid} R.~M.,  {Kuntschner} H.,  {Maraston} C.,
  {Conroy} C.,  2017, preprint, \href
  {http://adsabs.harvard.edu/abs/2017arXiv170909300B} {} (\mn@eprint {arXiv}
  {1709.09300})

\bibitem[\protect\citeauthoryear{{Bamford}, {Rojas}, {Nichol}, {Miller},
  {Wasserman}, {Genovese}  \& {Freeman}}{{Bamford} et~al.}{2008}]{bamford08}
{Bamford} S.~P.,  {Rojas} A.~L.,  {Nichol} R.~C.,  {Miller} C.~J.,  {Wasserman}
  L.,  {Genovese} C.~R.,   {Freeman} P.~E.,  2008, \mn@doi [\mnras]
  {10.1111/j.1365-2966.2008.13963.x}, \href
  {http://adsabs.harvard.edu/abs/2008MNRAS.391..607B} {391, 607}

\bibitem[\protect\citeauthoryear{{Bruzual} \& {Charlot}}{{Bruzual} \&
  {Charlot}}{2003}]{BC03}
{Bruzual} G.,  {Charlot} S.,  2003, \mn@doi [\mnras]
  {10.1046/j.1365-8711.2003.06897.x}, \href
  {http://adsabs.harvard.edu/abs/2003MNRAS.344.1000B} {344, 1000}

\bibitem[\protect\citeauthoryear{{Caccianiga}, {March{\~a}}, {Ant{\'o}n},
  {Mack}  \& {Neeser}}{{Caccianiga} et~al.}{2002}]{caccianiga2002}
{Caccianiga} A.,  {March{\~a}} M.~J.,  {Ant{\'o}n} S.,  {Mack} K.-H.,
  {Neeser} M.~J.,  2002, \mn@doi [\mnras] {10.1046/j.1365-8711.2002.05062.x},
  \href {http://adsabs.harvard.edu/abs/2002MNRAS.329..877C} {329, 877}

\bibitem[\protect\citeauthoryear{{Capozzi}, {Maraston}, {Daddi}, {Renzini},
  {Strazzullo}  \& {Gobat}}{{Capozzi} et~al.}{2016}]{capozzi16}
{Capozzi} D.,  {Maraston} C.,  {Daddi} E.,  {Renzini} A.,  {Strazzullo} V.,
  {Gobat} R.,  2016, \mn@doi [\mnras] {10.1093/mnras/stv2692}, \href
  {http://adsabs.harvard.edu/abs/2016MNRAS.456..790C} {456, 790}

\bibitem[\protect\citeauthoryear{{Cardelli}, {Clayton}  \& {Mathis}}{{Cardelli}
  et~al.}{1989}]{ccm}
{Cardelli} J.~A.,  {Clayton} G.~C.,   {Mathis} J.~S.,  1989, \mn@doi [\apj]
  {10.1086/167900}, \href {http://adsabs.harvard.edu/abs/1989ApJ...345..245C}
  {345, 245}

\bibitem[\protect\citeauthoryear{{Cassisi}, {degl'Innocenti}  \&
  {Salaris}}{{Cassisi} et~al.}{1997a}]{cassisi97b}
{Cassisi} S.,  {degl'Innocenti} S.,   {Salaris} M.,  1997a, \mn@doi [\mnras]
  {10.1093/mnras/290.3.515}, \href
  {http://adsabs.harvard.edu/abs/1997MNRAS.290..515C} {290, 515}

\bibitem[\protect\citeauthoryear{{Cassisi}, {Castellani}  \&
  {Castellani}}{{Cassisi} et~al.}{1997b}]{cassisi97a}
{Cassisi} S.,  {Castellani} M.,   {Castellani} V.,  1997b, \aap, \href
  {http://adsabs.harvard.edu/abs/1997A%26A...317..108C} {317, 108}

\bibitem[\protect\citeauthoryear{{Chen}, {Liang}, {Hammer}, {Prugniel},
  {Zhong}, {Rodrigues}, {Zhao}  \& {Flores}}{{Chen} et~al.}{2010}]{chen10}
{Chen} X.~Y.,  {Liang} Y.~C.,  {Hammer} F.,  {Prugniel} P.,  {Zhong} G.~H.,
  {Rodrigues} M.,  {Zhao} Y.~H.,   {Flores} H.,  2010, \mn@doi [\aap]
  {10.1051/0004-6361/200913894}, \href
  {http://adsabs.harvard.edu/abs/2010A%26A...515A.101C} {515, A101}

\bibitem[\protect\citeauthoryear{{Cid Fernandes}, {Gu}, {Melnick}, {Terlevich},
  {Terlevich}, {Kunth}, {Rodrigues Lacerda}  \& {Joguet}}{{Cid Fernandes}
  et~al.}{2004}]{CF04}
{Cid Fernandes} R.,  {Gu} Q.,  {Melnick} J.,  {Terlevich} E.,  {Terlevich} R.,
  {Kunth} D.,  {Rodrigues Lacerda} R.,   {Joguet} B.,  2004, \mn@doi [\mnras]
  {10.1111/j.1365-2966.2004.08321.x}, \href
  {http://adsabs.harvard.edu/abs/2004MNRAS.355..273C} {355, 273}

\bibitem[\protect\citeauthoryear{{Cid Fernandes}, {Mateus}, {Sodr{\'e}},
  {Stasi{\'n}ska}  \& {Gomes}}{{Cid Fernandes} et~al.}{2005}]{CF05}
{Cid Fernandes} R.,  {Mateus} A.,  {Sodr{\'e}} L.,  {Stasi{\'n}ska} G.,
  {Gomes} J.~M.,  2005, \mn@doi [\mnras] {10.1111/j.1365-2966.2005.08752.x},
  \href {http://adsabs.harvard.edu/abs/2005MNRAS.358..363C} {358, 363}

\bibitem[\protect\citeauthoryear{{Conroy}}{{Conroy}}{2013}]{conroy13}
{Conroy} C.,  2013, \mn@doi [\araa] {10.1146/annurev-astro-082812-141017},
  \href {http://adsabs.harvard.edu/abs/2013ARA%26A..51..393C} {51, 393}

\bibitem[\protect\citeauthoryear{{Conroy} \& {Gunn}}{{Conroy} \&
  {Gunn}}{2010}]{conroyEgunn10}
{Conroy} C.,  {Gunn} J.~E.,  2010, \mn@doi [\apj]
  {10.1088/0004-637X/712/2/833}, \href
  {http://adsabs.harvard.edu/abs/2010ApJ...712..833C} {712, 833}

\bibitem[\protect\citeauthoryear{{Conroy}, {Gunn}  \& {White}}{{Conroy}
  et~al.}{2009}]{C09}
{Conroy} C.,  {Gunn} J.~E.,   {White} M.,  2009, \mn@doi [\apj]
  {10.1088/0004-637X/699/1/486}, \href
  {http://adsabs.harvard.edu/abs/2009ApJ...699..486C} {699, 486}

\bibitem[\protect\citeauthoryear{{Cushing}, {Vacca}  \& {Rayner}}{{Cushing}
  et~al.}{2004}]{cushing04}
{Cushing} M.~C.,  {Vacca} W.~D.,   {Rayner} J.~T.,  2004, \mn@doi [\pasp]
  {10.1086/382907}, \href {http://adsabs.harvard.edu/abs/2004PASP..116..362C}
  {116, 362}

\bibitem[\protect\citeauthoryear{{Cushing}, {Rayner}  \& {Vacca}}{{Cushing}
  et~al.}{2005}]{cushing05}
{Cushing} M.~C.,  {Rayner} J.~T.,   {Vacca} W.~D.,  2005, \mn@doi [\apj]
  {10.1086/428040}, \href {http://adsabs.harvard.edu/abs/2005ApJ...623.1115C}
  {623, 1115}

\bibitem[\protect\citeauthoryear{{Dametto}, {Riffel}, {Pastoriza},
  {Rodr{\'{\i}}guez-Ardila}, {Hernandez-Jimenez}  \& {Carvalho}}{{Dametto}
  et~al.}{2014}]{dametto14}
{Dametto} N.~Z.,  {Riffel} R.,  {Pastoriza} M.~G.,  {Rodr{\'{\i}}guez-Ardila}
  A.,  {Hernandez-Jimenez} J.~A.,   {Carvalho} E.~A.,  2014, \mn@doi [\mnras]
  {10.1093/mnras/stu1243}, \href
  {http://adsabs.harvard.edu/abs/2014MNRAS.443.1754D} {443, 1754}

\bibitem[\protect\citeauthoryear{{Dressel} \& {Condon}}{{Dressel} \&
  {Condon}}{1978}]{dresselcondon78}
{Dressel} L.~L.,  {Condon} J.~J.,  1978, \mn@doi [\apjs] {10.1086/190491},
  \href {http://adsabs.harvard.edu/abs/1978ApJS...36...53D} {36, 53}

\bibitem[\protect\citeauthoryear{{Forman}, {Jones}  \& {Tucker}}{{Forman}
  et~al.}{1985}]{forman85}
{Forman} W.,  {Jones} C.,   {Tucker} W.,  1985, \mn@doi [\apj]
  {10.1086/163218}, \href {http://adsabs.harvard.edu/abs/1985ApJ...293..102F}
  {293, 102}

\bibitem[\protect\citeauthoryear{{Gomes} et~al.,}{{Gomes}
  et~al.}{2016}]{gomes16}
{Gomes} J.~M.,  et~al., 2016, \mn@doi [\aap] {10.1051/0004-6361/201525976},
  \href {http://adsabs.harvard.edu/abs/2016A%26A...588A..68G} {588, A68}

\bibitem[\protect\citeauthoryear{{H{\"o}fner}, {Loidl}, {Aringer},
  {J{\o}rgensen}  \& {Hron}}{{H{\"o}fner} et~al.}{2000}]{hofner+00}
{H{\"o}fner} S.,  {Loidl} R.,  {Aringer} B.,  {J{\o}rgensen} U.~G.,   {Hron}
  J.,  2000, in {Salama} A.,  {Kessler} M.~F.,  {Leech} K.,   {Schulz} B.,
  eds,  ESA Special Publication Vol. 456, ISO Beyond the Peaks: The 2nd ISO
  Workshop on Analytical Spectroscopy. p.~299

\bibitem[\protect\citeauthoryear{{Jones}, {Forman}, {Vikhlinin}, {Markevitch},
  {David}, {Warmflash}, {Murray}  \& {Nulsen}}{{Jones} et~al.}{2002}]{jones02}
{Jones} C.,  {Forman} W.,  {Vikhlinin} A.,  {Markevitch} M.,  {David} L.,
  {Warmflash} A.,  {Murray} S.,   {Nulsen} P.~E.~J.,  2002, \mn@doi [\apjl]
  {10.1086/340114}, \href {http://adsabs.harvard.edu/abs/2002ApJ...567L.115J}
  {567, L115}

\bibitem[\protect\citeauthoryear{{Kehrig} et~al.,}{{Kehrig}
  et~al.}{2012}]{kehrig12}
{Kehrig} C.,  et~al., 2012, \mn@doi [\aap] {10.1051/0004-6361/201118357}, \href
  {http://adsabs.harvard.edu/abs/2012A%26A...540A..11K} {540, A11}

\bibitem[\protect\citeauthoryear{{Kriek} et~al.,}{{Kriek}
  et~al.}{2010}]{kriek+10}
{Kriek} M.,  et~al., 2010, \mn@doi [\apjl] {10.1088/2041-8205/722/1/L64}, \href
  {http://adsabs.harvard.edu/abs/2010ApJ...722L..64K} {722, L64}

\bibitem[\protect\citeauthoryear{{Kroupa}}{{Kroupa}}{2001}]{kroupa01}
{Kroupa} P.,  2001, \mn@doi [\mnras] {10.1046/j.1365-8711.2001.04022.x}, \href
  {http://adsabs.harvard.edu/abs/2001MNRAS.322..231K} {322, 231}

\bibitem[\protect\citeauthoryear{{Lan{\c c}on} \& {Mouhcine}}{{Lan{\c c}on} \&
  {Mouhcine}}{2002}]{lanconEmouhcine02}
{Lan{\c c}on} A.,  {Mouhcine} M.,  2002, \mn@doi [\aap]
  {10.1051/0004-6361:20020585}, \href
  {http://adsabs.harvard.edu/abs/2002A%26A...393..167L} {393, 167}

\bibitem[\protect\citeauthoryear{{Le Borgne} et~al.,}{{Le Borgne}
  et~al.}{2003}]{lebogne03}
{Le Borgne} J.-F.,  et~al., 2003, \mn@doi [\aap] {10.1051/0004-6361:20030243},
  \href {http://adsabs.harvard.edu/abs/2003A%26A...402..433L} {402, 433}

\bibitem[\protect\citeauthoryear{{Lejeune}, {Cuisinier}  \& {Buser}}{{Lejeune}
  et~al.}{1997}]{lejeune97}
{Lejeune} T.,  {Cuisinier} F.,   {Buser} R.,  1997, \mn@doi [\aaps]
  {10.1051/aas:1997373}, \href
  {http://adsabs.harvard.edu/abs/1997A%26AS..125..229L} {125}

\bibitem[\protect\citeauthoryear{{Lejeune}, {Cuisinier}  \& {Buser}}{{Lejeune}
  et~al.}{1998}]{lejeune98}
{Lejeune} T.,  {Cuisinier} F.,   {Buser} R.,  1998, \mn@doi [\aaps]
  {10.1051/aas:1998405}, \href
  {http://adsabs.harvard.edu/abs/1998A%26AS..130...65L} {130, 65}

\bibitem[\protect\citeauthoryear{{Maraston}}{{Maraston}}{2005}]{M05}
{Maraston} C.,  2005, \mn@doi [\mnras] {10.1111/j.1365-2966.2005.09270.x},
  \href {http://adsabs.harvard.edu/abs/2005MNRAS.362..799M} {362, 799}

\bibitem[\protect\citeauthoryear{{Maraston} \& {Str{\"o}mb{\"a}ck}}{{Maraston}
  \& {Str{\"o}mb{\"a}ck}}{2011}]{M11}
{Maraston} C.,  {Str{\"o}mb{\"a}ck} G.,  2011, \mn@doi [\mnras]
  {10.1111/j.1365-2966.2011.19738.x}, \href
  {http://adsabs.harvard.edu/abs/2011MNRAS.418.2785M} {418, 2785}

\bibitem[\protect\citeauthoryear{{Marigo}, {Girardi}, {Bressan}, {Groenewegen},
  {Silva}  \& {Granato}}{{Marigo} et~al.}{2008}]{marigo08}
{Marigo} P.,  {Girardi} L.,  {Bressan} A.,  {Groenewegen} M.~A.~T.,  {Silva}
  L.,   {Granato} G.~L.,  2008, \mn@doi [\aap] {10.1051/0004-6361:20078467},
  \href {http://adsabs.harvard.edu/abs/2008A%26A...482..883M} {482, 883}

\bibitem[\protect\citeauthoryear{{Martel} et~al.,}{{Martel}
  et~al.}{2004}]{martel2004}
{Martel} A.~R.,  et~al., 2004, \mn@doi [\aj] {10.1086/425628}, \href
  {http://adsabs.harvard.edu/abs/2004AJ....128.2758M} {128, 2758}

\bibitem[\protect\citeauthoryear{{Martins}, {Lanfranchi}, {Gon{\c c}alves},
  {Magrini}, {Teodorescu}  \& {Quireza}}{{Martins} et~al.}{2012}]{martins12}
{Martins} L.~P.,  {Lanfranchi} G.,  {Gon{\c c}alves} D.~R.,  {Magrini} L.,
  {Teodorescu} A.~M.,   {Quireza} C.,  2012, \mn@doi [\mnras]
  {10.1111/j.1365-2966.2011.19954.x}, \href
  {http://adsabs.harvard.edu/abs/2012MNRAS.419.3159M} {419, 3159}

\bibitem[\protect\citeauthoryear{{Martins}, {Rodr{\'{\i}}guez-Ardila}, {Diniz},
  {Gruenwald}  \& {de Souza}}{{Martins} et~al.}{2013}]{martins13b}
{Martins} L.~P.,  {Rodr{\'{\i}}guez-Ardila} A.,  {Diniz} S.,  {Gruenwald} R.,
  {de Souza} R.,  2013, \mn@doi [\mnras] {10.1093/mnras/stt296}, \href
  {http://adsabs.harvard.edu/abs/2013MNRAS.431.1823M} {431, 1823}

\bibitem[\protect\citeauthoryear{{Meneses-Goytia}, {Peletier}, {Trager}  \&
  {Vazdekis}}{{Meneses-Goytia} et~al.}{2015}]{MG15}
{Meneses-Goytia} S.,  {Peletier} R.~F.,  {Trager} S.~C.,   {Vazdekis} A.,
  2015, \mn@doi [\aap] {10.1051/0004-6361/201423838}, \href
  {http://adsabs.harvard.edu/abs/2015A%26A...582A..97M} {582, A97}

\bibitem[\protect\citeauthoryear{{Nilson}}{{Nilson}}{1973}]{nilson73}
{Nilson} P.,  1973, {Uppsala general catalogue of galaxies}

\bibitem[\protect\citeauthoryear{{No{\"e}l}, {Greggio}, {Renzini}, {Carollo}
  \& {Maraston}}{{No{\"e}l} et~al.}{2013}]{noel+13}
{No{\"e}l} N.~E.~D.,  {Greggio} L.,  {Renzini} A.,  {Carollo} C.~M.,
  {Maraston} C.,  2013, \mn@doi [\apj] {10.1088/0004-637X/772/1/58}, \href
  {http://adsabs.harvard.edu/abs/2013ApJ...772...58N} {772, 58}

\bibitem[\protect\citeauthoryear{{Padilla} \& {Strauss}}{{Padilla} \&
  {Strauss}}{2008}]{PadillaStrauss08}
{Padilla} N.~D.,  {Strauss} M.~A.,  2008, \mn@doi [\mnras]
  {10.1111/j.1365-2966.2008.13480.x}, \href
  {http://adsabs.harvard.edu/abs/2008MNRAS.388.1321P} {388, 1321}

\bibitem[\protect\citeauthoryear{{Pandge}, {Vagshette}, {David}  \&
  {Patil}}{{Pandge} et~al.}{2012}]{pandge12}
{Pandge} M.~B.,  {Vagshette} N.~D.,  {David} L.~P.,   {Patil} M.~K.,  2012,
  \mn@doi [\mnras] {10.1111/j.1365-2966.2011.20358.x}, \href
  {http://adsabs.harvard.edu/abs/2012MNRAS.421..808P} {421, 808}

\bibitem[\protect\citeauthoryear{{Pietrinferni}, {Cassisi}, {Salaris}  \&
  {Castelli}}{{Pietrinferni} et~al.}{2004}]{BaSTI}
{Pietrinferni} A.,  {Cassisi} S.,  {Salaris} M.,   {Castelli} F.,  2004,
  \mn@doi [\apj] {10.1086/422498}, \href
  {http://adsabs.harvard.edu/abs/2004ApJ...612..168P} {612, 168}

\bibitem[\protect\citeauthoryear{{Raichur}, {Das}, {Herrero}, {Shastri}  \&
  {Kantharia}}{{Raichur} et~al.}{2015}]{raichur15}
{Raichur} H.,  {Das} M.,  {Herrero} A.~A.,  {Shastri} P.,   {Kantharia} N.~G.,
  2015, \mn@doi [\apss] {10.1007/s10509-015-2290-y}, \href
  {http://adsabs.harvard.edu/abs/2015Ap%26SS.357...32R} {357, 32}

\bibitem[\protect\citeauthoryear{{Rayner}, {Cushing}  \& {Vacca}}{{Rayner}
  et~al.}{2009}]{rayner09}
{Rayner} J.~T.,  {Cushing} M.~C.,   {Vacca} W.~D.,  2009, \mn@doi [\apjs]
  {10.1088/0067-0049/185/2/289}, \href
  {http://adsabs.harvard.edu/abs/2009ApJS..185..289R} {185, 289}

\bibitem[\protect\citeauthoryear{{Rickes}, {Pastoriza}  \& {Bonatto}}{{Rickes}
  et~al.}{2009}]{rickes09}
{Rickes} M.~G.,  {Pastoriza} M.~G.,   {Bonatto} C.,  2009, \mn@doi [\aap]
  {10.1051/0004-6361/200811500}, \href
  {http://adsabs.harvard.edu/abs/2009A%26A...505...73R} {505, 73}

\bibitem[\protect\citeauthoryear{{Riffel} \& {Borges Vale}}{{Riffel} \& {Borges
  Vale}}{2011}]{pacce}
{Riffel} R.,  {Borges Vale} T.,  2011, \mn@doi [\apss]
  {10.1007/s10509-011-0731-9}, \href
  {http://adsabs.harvard.edu/abs/2011Ap%26SS.334..351R} {334, 351}

\bibitem[\protect\citeauthoryear{{Riffel}, {Pastoriza},
  {Rodr{\'{\i}}guez-Ardila}  \& {Maraston}}{{Riffel} et~al.}{2007}]{riffel07}
{Riffel} R.,  {Pastoriza} M.~G.,  {Rodr{\'{\i}}guez-Ardila} A.,   {Maraston}
  C.,  2007, \mn@doi [\apjl] {10.1086/517999}, \href
  {http://adsabs.harvard.edu/abs/2007ApJ...659L.103R} {659, L103}

\bibitem[\protect\citeauthoryear{{Riffel}, {Pastoriza},
  {Rodr{\'{\i}}guez-Ardila}  \& {Maraston}}{{Riffel} et~al.}{2008}]{riffel08}
{Riffel} R.,  {Pastoriza} M.~G.,  {Rodr{\'{\i}}guez-Ardila} A.,   {Maraston}
  C.,  2008, \mn@doi [\mnras] {10.1111/j.1365-2966.2008.13440.x}, \href
  {http://adsabs.harvard.edu/abs/2008MNRAS.388..803R} {388, 803}

\bibitem[\protect\citeauthoryear{{Riffel}, {Pastoriza},
  {Rodr{\'{\i}}guez-Ardila}  \& {Bonatto}}{{Riffel} et~al.}{2009}]{riffel09}
{Riffel} R.,  {Pastoriza} M.~G.,  {Rodr{\'{\i}}guez-Ardila} A.,   {Bonatto} C.,
   2009, \mn@doi [\mnras] {10.1111/j.1365-2966.2009.15448.x}, \href
  {http://adsabs.harvard.edu/abs/2009MNRAS.400..273R} {400, 273}

\bibitem[\protect\citeauthoryear{{Riffel}, {Riffel}, {Ferrari}  \&
  {Storchi-Bergmann}}{{Riffel} et~al.}{2011}]{riffel11}
{Riffel} R.,  {Riffel} R.~A.,  {Ferrari} F.,   {Storchi-Bergmann} T.,  2011,
  \mn@doi [\mnras] {10.1111/j.1365-2966.2011.19061.x}, \href
  {http://adsabs.harvard.edu/abs/2011MNRAS.416..493R} {416, 493}

\bibitem[\protect\citeauthoryear{{Riffel} et~al.,}{{Riffel}
  et~al.}{2015}]{riffel15}
{Riffel} R.,  et~al., 2015, \mn@doi [\mnras] {10.1093/mnras/stv866}, \href
  {http://adsabs.harvard.edu/abs/2015MNRAS.450.3069R} {450, 3069}

\bibitem[\protect\citeauthoryear{{R{\"o}ck}, {Vazdekis}, {Ricciardelli},
  {Peletier}, {Knapen}  \& {Falc{\'o}n-Barroso}}{{R{\"o}ck}
  et~al.}{2016}]{rock16}
{R{\"o}ck} B.,  {Vazdekis} A.,  {Ricciardelli} E.,  {Peletier} R.~F.,  {Knapen}
  J.~H.,   {Falc{\'o}n-Barroso} J.,  2016, \mn@doi [\aap]
  {10.1051/0004-6361/201527570}, \href
  {http://adsabs.harvard.edu/abs/2016A%26A...589A..73R} {589, A73}

\bibitem[\protect\citeauthoryear{{Salpeter}}{{Salpeter}}{1955}]{salpeter}
{Salpeter} E.~E.,  1955, \mn@doi [\apj] {10.1086/145971}, \href
  {http://adsabs.harvard.edu/abs/1955ApJ...121..161S} {121, 161}

\bibitem[\protect\citeauthoryear{{S{\'a}nchez} et~al.,}{{S{\'a}nchez}
  et~al.}{2012}]{sanchez+12}
{S{\'a}nchez} S.~F.,  et~al., 2012, \mn@doi [\aap]
  {10.1051/0004-6361/201117353}, \href
  {http://adsabs.harvard.edu/abs/2012A%26A...538A...8S} {538, A8}

\bibitem[\protect\citeauthoryear{{S{\'a}nchez} et~al.,}{{S{\'a}nchez}
  et~al.}{2016}]{sanchez+16}
{S{\'a}nchez} S.~F.,  et~al., 2016, \mn@doi [\aap]
  {10.1051/0004-6361/201628661}, \href
  {http://adsabs.harvard.edu/abs/2016A%26A...594A..36S} {594, A36}

\bibitem[\protect\citeauthoryear{{Schaller}, {Schaerer}, {Meynet}  \&
  {Maeder}}{{Schaller} et~al.}{1992}]{schaller92}
{Schaller} G.,  {Schaerer} D.,  {Meynet} G.,   {Maeder} A.,  1992, \aaps, \href
  {http://adsabs.harvard.edu/abs/1992A%26AS...96..269S} {96, 269}

\bibitem[\protect\citeauthoryear{{Schmidt}, {Copetti}, {Alloin}  \&
  {Jablonka}}{{Schmidt} et~al.}{1991}]{schmidt91}
{Schmidt} A.~A.,  {Copetti} M.~V.~F.,  {Alloin} D.,   {Jablonka} P.,  1991,
  \mn@doi [\mnras] {10.1093/mnras/249.4.766}, \href
  {http://adsabs.harvard.edu/abs/1991MNRAS.249..766S} {249, 766}

\bibitem[\protect\citeauthoryear{{Stanger} \& {Warwick}}{{Stanger} \&
  {Warwick}}{1986}]{stangerwarwick86}
{Stanger} V.~J.,  {Warwick} R.~S.,  1986, \mn@doi [\mnras]
  {10.1093/mnras/220.2.363}, \href
  {http://adsabs.harvard.edu/abs/1986MNRAS.220..363S} {220, 363}

\bibitem[\protect\citeauthoryear{{Vacca}, {Cushing}  \& {Rayner}}{{Vacca}
  et~al.}{2004}]{vacca04}
{Vacca} W.~D.,  {Cushing} M.~C.,   {Rayner} J.~T.,  2004, \mn@doi [\pasp]
  {10.1086/382906}, \href {http://adsabs.harvard.edu/abs/2004PASP..116..352V}
  {116, 352}

\bibitem[\protect\citeauthoryear{{V{\'e}ron-Cetty} \&
  {V{\'e}ron}}{{V{\'e}ron-Cetty} \& {V{\'e}ron}}{2006}]{veroncetty06}
{V{\'e}ron-Cetty} M.-P.,  {V{\'e}ron} P.,  2006, \mn@doi [\aap]
  {10.1051/0004-6361:20065177}, \href
  {http://adsabs.harvard.edu/abs/2006A%26A...455..773V} {455, 773}

\bibitem[\protect\citeauthoryear{{Walcher} et~al.,}{{Walcher}
  et~al.}{2014}]{husemann+14}
{Walcher} C.~J.,  et~al., 2014, \mn@doi [\aap] {10.1051/0004-6361/201424198},
  \href {http://adsabs.harvard.edu/abs/2014A%26A...569A...1W} {569, A1}

\bibitem[\protect\citeauthoryear{{Westera}, {Lejeune}, {Buser}, {Cuisinier}  \&
  {Bruzual}}{{Westera} et~al.}{2002}]{westera02}
{Westera} P.,  {Lejeune} T.,  {Buser} R.,  {Cuisinier} F.,   {Bruzual} G.,
  2002, \mn@doi [\aap] {10.1051/0004-6361:20011493}, \href
  {http://adsabs.harvard.edu/abs/2002A%26A...381..524W} {381, 524}

\bibitem[\protect\citeauthoryear{{Wilson} et~al.,}{{Wilson}
  et~al.}{2004}]{wilson04}
{Wilson} J.~C.,  et~al., 2004, in {Moorwood} A.~F.~M.,  {Iye} M.,  eds,
  \procspie Vol. 5492, Ground-based Instrumentation for Astronomy. pp
  1295--1305, \mn@doi{10.1117/12.550925}

\bibitem[\protect\citeauthoryear{{Worthey}}{{Worthey}}{1994}]{worthey94}
{Worthey} G.,  1994, \mn@doi [\apjs] {10.1086/192096}, \href
  {http://adsabs.harvard.edu/abs/1994ApJS...95..107W} {95, 107}

\bibitem[\protect\citeauthoryear{{Wrobel}}{{Wrobel}}{1984}]{wrobel84}
{Wrobel} J.~M.,  1984, \mn@doi [\apj] {10.1086/162436}, \href
  {http://adsabs.harvard.edu/abs/1984ApJ...284..531W} {284, 531}

\bibitem[\protect\citeauthoryear{{Zibetti}, {Gallazzi}, {Charlot}, {Pierini}
  \& {Pasquali}}{{Zibetti} et~al.}{2013}]{zibetti13}
{Zibetti} S.,  {Gallazzi} A.,  {Charlot} S.,  {Pierini} D.,   {Pasquali} A.,
  2013, \mn@doi [\mnras] {10.1093/mnras/sts126}, \href
  {http://adsabs.harvard.edu/abs/2013MNRAS.428.1479Z} {428, 1479}

\bibitem[\protect\citeauthoryear{{de Vaucouleurs}, {de Vaucouleurs}, {Corwin},
  {Buta}, {Paturel}  \& {Fouqu{\'e}}}{{de Vaucouleurs}
  et~al.}{1991}]{devaucouleurs91}
{de Vaucouleurs} G.,  {de Vaucouleurs} A.,  {Corwin} Jr. H.~G.,  {Buta} R.~J.,
  {Paturel} G.,   {Fouqu{\'e}} P.,  1991, {Third Reference Catalogue of Bright
  Galaxies. Volume I: Explanations and references. Volume II: Data for galaxies
  between 0$^{h}$ and 12$^{h}$. Volume III: Data for galaxies between 12$^{h}$
  and 24$^{h}$.}

\makeatother
\end{thebibliography}


\appendix
\section{Individual description of the sample}
\label{sec:sample}

\subsection{NGC 4636}

NGC 4636 is a giant elliptical galaxy with LINER emission. Is one of the nearest and more X-ray luminous ellipticals, with $L_X ~ 2x10^41$ ergs/s \citep{jones02}. It is surrounded by an extended corona of hot gas \citep{forman85}, and has an asymmetric gas distribution, probably the result of irregular flows \citep{stangerwarwick86}.

\subsection{NGC 5905}

This galaxy is one of a few that did not have a previously evidence of an AGN but that had X-ray eruptions observed, what happened in 1991 and 1992 with Chandra telescope. Later, X-ray flux started to decrease in a rate consistent with the expected for a Tidal Disruption Event (TDE). Recent observations showed that the infrared flux is dominated by star formation, what suggests that the radio emission is caused by circumnuclear star formation. Besides, no radio emission consistent with the TDE event was detected \citep{raichur15}. Among the galaxies of our sample, it is the only that has clearly visible NIR emission lines an the only non-ETG.

\subsection{NGC 5966}
NGC 5966 is an elliptical galaxy with a faint bar-like feature centered at the nucleus. Stars and gas are decoupled, with the gas showing an elongated emission feature, in such a way that an ionization cone or a decoupled rotational disk are two possible interpretations \citep{kehrig12}. According to these authors, the diagnostic diagrams indicate the presence of a LINER nucleus and a LINER-like gas emission extending $\sim$5~kpc outward from the nucleus, also LINER-like. The presence of a nuclear ionizing source seems to be required to shape the elongated gas emission feature in the ionization cone'' scenario, although ionization by pAGB stars cannot be ruled out. On the other hand, \citet{gomes16} classified this object as type i ETG, also reporting that the absence of ongoing star formation throughout the galaxy lends support to the idea that its gas outflow is powered by an AGN hosted in its LINER nucleus. According to \citet{gomes16}, a Type i ETG is a system with a nearly constant EW(H$\alpha$) in their extranuclear component, compatible with the hypothesis of photoionization by pAGB as the main driver of extended warm interstellar medium (WIM) emission.

\subsection{NGC 6081}
It is a galaxy with radio emission \citep{dresselcondon78}. Also, \citet{gomes16} classified it as a type i ETG, reporting that sources other than pAGB stars dominate in less than 7\% of the area of the galaxy. Its SDSS spectra is dominated by stellar population, with strong absorption bands and [OIII], [NI], $H\alpha$ and [SII] typical of LINERs. It's APO spectrum shows a stellar continuum dominated source with strong CO bands and absence of emission lines.

\subsection{NGC 6146}
NGC 6146 is an elliptical galaxy with a radio jet \citep{wrobel84}. \citet{caccianiga2002} reported the detection of CaII H and K, G, Mg Ib and Na Id absorption bands and a weak $H\alpha$ emission line. \citet{gomes16} classified this galaxy as type i ETG, also reporting that the absence of star formation and the suggestion of an outflow are compatible with the idea of a low luminosity AGN powering the nucleus. Its APO spectra shows a stellar continuum dominated source with strong CO bands and absence of emission lines.

\subsection{NGC 6338}
This galaxy is the brightest on Draco constellation. It posesses diffuse and ionized gas and dust filaments on kiloparsec scales \citep{martel2004}. X-ray observations indicate two or possibly three emission cavities of ellipsoidal shape with lower X-ray surface flux. They also found cold filamentary structures matching the $H \alpha$ emission and with high extinction regions seen on the optical extinction maps. This indicates a cooling mechanism generated by dust \citep{pandge12}. On the same work, they reported that a harder ionizing source is required to maintain such a high degree of ionization and that most of the jet power is mechanical. \citet{gomes16}, using CALIFA datacubes, classified this object as type ii ETG. 

\subsection{UGC 08234}
This galaxy is part of a galaxy group and posesses a spiral companion, UGC 08237 \citep{nilson73}. \citet{gomes16} classified this object as type ii ETG. There are no X-ray or radio properties reported in the literature. Its APO spectrum is dominated by stellar population, characterized by intense absorption bands and absence of emission lines.

\section{Emission Line Fluxes}
\label{sec:EmissionLines}



\begin{table}
\centering
\caption{NIR emission line fluxes measured for NGC\,5905.}
\label{tab:elfluxes}
\begin{tabular}{lcc}
\hline
Line     & Flux  & FWHM\\
         & ($\times 10^{-15}$ ergs/s/cm$^2$) & (km/s)\\
\hline
[FeII] $\lambda$12570$\AA$ & 9.33  & 806\\
Pa$\beta$                  & 6.72  & 315\\
H$_2$ $\lambda$21218$\AA$  & 2.68  & 385\\
Br$\gamma$                 & 3.5   & 462\\
\hline
\end{tabular}
\end{table}

\par

  In the optical region, according to \citet[BPT]{bpt}, the line ratios [O\,{\sc iii}]$\lambda$5007 \AA/H$\beta$ and [N\,{\sc ii}]$\lambda$6583\AA/H$\alpha$ can be used to classify the excitation mechanism that powers the optical lines based on the position of the points in the diagram. After running the stellar population synthesis for the SDSS spectra, we subtracted the stellar component and measured the emission line fluxes by fitting gaussians to the emission lines. Although there are still controversies on individual objects \citep[e.g.][{who found a galaxy with no signs of activity but whose line ratios put it as a Seyfert galaxy}]{martins12}, these diagrams can be used with great confidence to study the nature of the central source that powers the observed emission lines.

  \par

\begin{figure}
        \noindent
	\includegraphics[width=\linewidth]{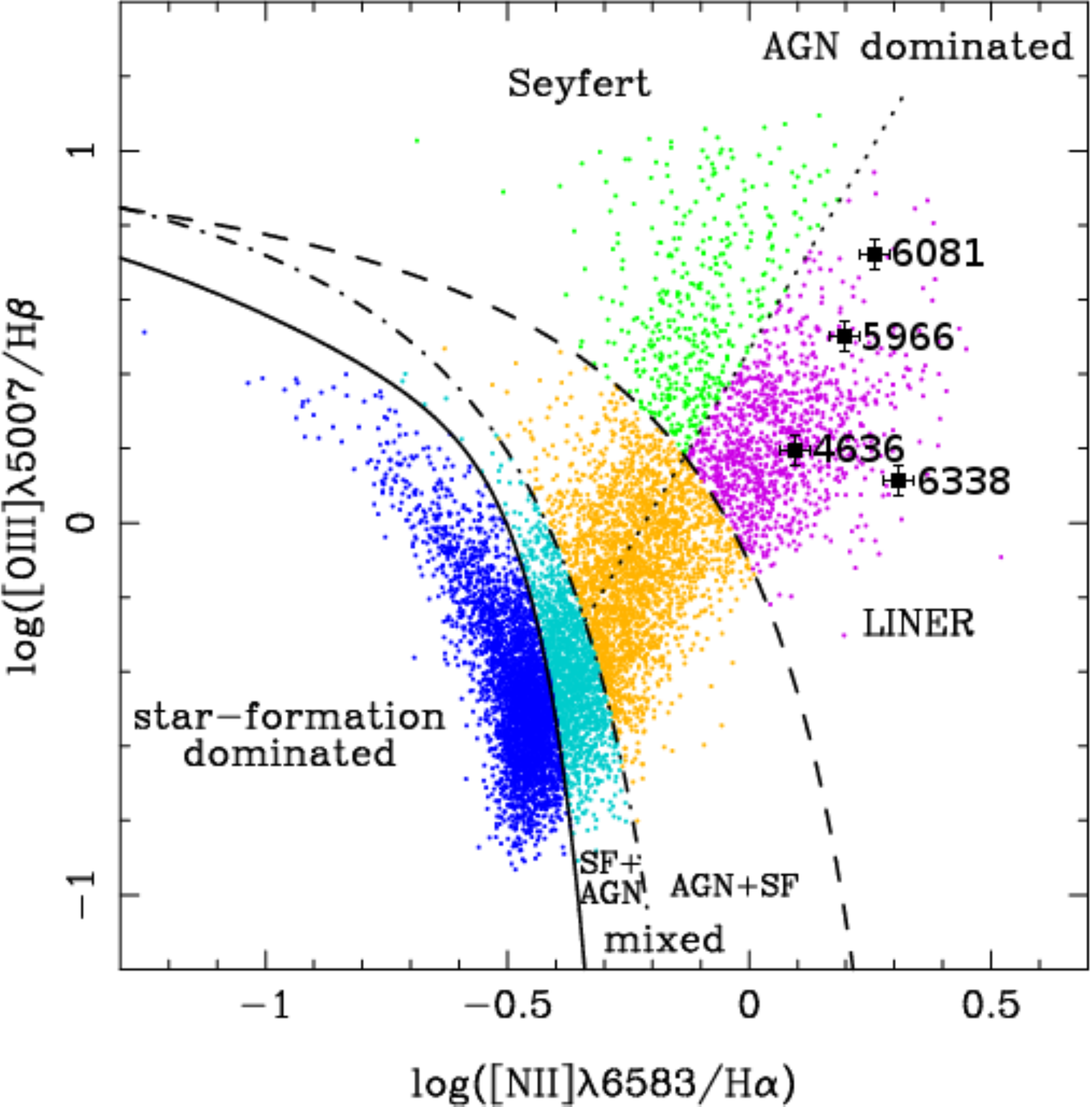}
    \caption{BPT diagram for the galaxies with optical spectra. The number of the NGC catalog is shown on the right side of each point. The points were plotted over the diagram from \citet{bamford08}}
    \label{fig:bpt}
\end{figure}

The emission line fluxes measured in the galaxy sample are listed in Table~\ref{tab:opticlines} and their ratios are plotted in Figure~\ref{fig:bpt}. From the locus of points occupied by our targets, we confirm the LINER nature of the galaxies. Only NGC\,4636 and NGC\,5966 had already been previously classified as LINER \citep{veroncetty06,kehrig12}. The results obtained allow us to state that our sample is composed of 5 ETGs with LINER emission, 1 spiral galaxy and 1 ETG with no data about its emission lines.

\begin{table}
\centering
\setlength{\tabcolsep}{3pt}
\caption{Optical emission line fluxes in units of 10$^{-15}$ ergs/s/cm$^2$/$\AA$.}
\label{tab:opticlines}
\begin{tabular}{lccccccr}
\hline
Object   & H$\beta$  &[OIII]     & [NII]     & H$\alpha$ & [NII]     & [SII]     & [SII]    \\
         & 4861$\AA$ & 5007$\AA$ & 6549$\AA$ & 6563$\AA$ & 6585$\AA$ & 6718$\AA$ & 6732$\AA$\\
\hline
NGC 4636 &1.9        &2.7        &3.43       &6.84       &8.77       &3.48       &3.10    \\
NGC 5966 &0.87       &2.26       &1.75       &2.69       &3.92       &2.79       &2.00    \\
NGC 6081 &1.05       &3.7        &7.26       &6.73       &13.5       &5.61       &3.71    \\
NGC 6338 &1.82       &2.38       &4.37       &5.43       &11.5       &3.99       &2.87    \\
\hline
\end{tabular}
\end{table}

\bsp	
\label{lastpage}
\end{document}